# Seasonal transmission potential and activity peaks of the new influenza A(H1N1): a Monte Carlo likelihood analysis based on human mobility

Duygu Balcan[1,2]*, Hao Hu[1-3]*, Bruno Goncalves[1,2]*, Paolo Bajardi[4,5]*, Chiara Poletto[4]*, Jose J Ramasco[4], Daniela Paolotti[4], Nicola Perra[1,6,7], Michele Tizzoni[4,8], Wouter Van den Broeck[4], Vittoria Colizza[4] and Alessandro Vespignani[1,2,4]§

[1]Center for Complex Networks and Systems Research, School of Informatics and Computing, Indiana University, Bloomington, IN, USA

[2]Pervasive Technology Institute, Indiana University, Bloomington, IN, USA

[3]Department of Physics, Indiana University, Bloomington, IN, USA
[4]Computational Epidemiology Laboratory, Institute for Scientific Interchange, Turin, Italy

[5]Centre de Physique Théorique, Université d'Aix-Marseille, Marseille , France

[6]Department of Physics, University of Cagliari, Cagliari, Italy

[7]Linkalab, Cagliari, Italy

[8] Scuola di Dottorato, Politecnico di Torino, Torino Italy

*These authors contributed equally to this work
§Corresponding author
Email addresses:

        DB: duygu.balcan@indiana.edu

        HH: hahu@indiana.edu

        BG: bgoncalv@indiana.edu

        PB: paolo.bajardi@isi.it

        CP: chiara.poletto@isi.it

        JJR: jramasco@isi.it

        DP: daniela.paolotti@isi.it

        NP: nperra@indiana.edu

        MT: michele.tizzoni@isi.it

        WVdB: wouter.vandenbroeck@isi.it

        VC: vcolizza@isi.it

        AV: alexv@indiana.edu

**Abstract**

**Background**

On 11 June the World Health Organization officially raised the phase of pandemic alert (with regard to the new H1N1 influenza strain) to level 6. As of 19 July, 137,232 cases of the H1N1 influenza strain have been officially confirmed in 142 different countries, and the pandemic unfolding in the Southern hemisphere is now under scrutiny to gain insights about the next winter wave in the Northern hemisphere. A major challenge is pre-empted by the need to estimate the transmission potential of the virus and to assess its dependence on seasonality aspects in order to be able to use numerical models capable of projecting the spatiotemporal pattern of the pandemic.

**Methods**

In the present work, we use a global structured metapopulation model integrating mobility and transportation data worldwide. The model considers data on 3,362 subpopulations in 220 different countries and individual mobility across them. The model generates stochastic realizations of the epidemic evolution worldwide considering 6 billion individuals, from which we can gather information such as prevalence, morbidity, number of secondary cases and number and date of imported cases for each subpopulation, all with a time resolution of 1 day. In order to estimate the transmission potential and the relevant model parameters we used the data on the chronology of the 2009 novel influenza A(H1N1). The method is based on the maximum likelihood analysis of the arrival time distribution generated by the model in 12 countries seeded by Mexico by using 1 million computationally simulated epidemics. An extended chronology including 93 countries worldwide seeded before 18 June was used to ascertain the seasonality effects.

**Results**

We found the best estimate $R_0$ = 1.75 (95% confidence interval (CI) 1.64 to 1.88) for the basic reproductive number. Correlation analysis allows the selection of the most probable seasonal behavior based on the observed pattern, leading to the identification of plausible scenarios for the future unfolding of the pandemic and the estimate of pandemic activity peaks in the different hemispheres. We provide estimates for the number of hospitalizations and the attack rate for the next wave as well as an extensive sensitivity analysis on the disease parameter values. We also studied the effect of systematic therapeutic use of antiviral drugs on the epidemic timeline.

**Conclusions**

The analysis shows the potential for an early epidemic peak occurring in October/November in the Northern hemisphere, likely before large-scale vaccination campaigns could be carried out. The baseline results refer to a worst-case scenario in which additional mitigation policies are not considered. We suggest that the planning of additional mitigation policies such as systematic antiviral treatments might be the key to delay the activity peak in order to restore the effectiveness of the vaccination programs.

**Background**

Estimating the transmission potential of a newly emerging virus is crucial when planning for adequate public health interventions to mitigate its spread and impact, and to forecast the expected epidemic scenarios through sophisticate computational approaches [1-4]. With the current outbreak of the new influenza A(H1N1) strain having reached pandemic proportions, the investigation of the influenza situation worldwide might provide the key to the understanding to the transmissibility observed in different regions and to the characterization of possible seasonal behavior. During the early phase of an outbreak, this task is hampered by inaccuracies and incompleteness of available information. Reporting is constrained by the difficulties in confirming large numbers of cases through specific tests and serological analysis. The cocirculation of multiple strains, the presence of asymptomatic cases that go undetected, the impossibility to monitor mild cases that do not seek health care and the possible delays in diagnosis and reporting, all worsen the situation. Early modeling approaches and statistical analysis show that the number of confirmed cases by the Mexican authorities during the early phase was underestimated by a factor ranging from one order of magnitude [5] to almost three [6]. The Centers for Disease Control (CDC) in the US estimate a 5% to 10% case detection, similar to other countries facing large outbreaks, with expected heterogeneities due to different surveillance systems. Even within the same country, the setup of enhanced monitoring led to improved notification with respect to the earlier phase, later relaxed as reporting requirements changed [7].

By contrast, the effort put in place by the World Health Organization (WHO) and health protection agencies worldwide is providing an unprecedented amount of data and, at last, the possibility of following in real time the pandemic chronology on the global scale. In particular, the border controls and the enhanced surveillance aimed at detecting the first cases reaching uninfected countries appear to provide more reliable and timely information with respect to the raw count of cases as local transmission occurs, and this data has already been used for early assessment of the number of cases in Mexico [5]. Moreover, data on international passenger flows from Mexico was found to display a strong correlation with confirmed H1N1 importations from Mexico [8]. Here we present an estimate of the reproduction number, $R_0$, (that is, the average number of secondary cases produced by a primary case [9]) of the current H1N1 epidemic based on knowledge of human mobility patterns. We use the GLEaM (for GLobal Epidemic and Mobility) structured metapopulation model [10] for the worldwide evolution of the pandemic and perform a maximum likelihood analysis of the parameters against the actual chronology of newly infected countries. The method is computationally intensive as it involves a Monte Carlo generation of the distribution of arrival time of the infection in each country based on the analysis of $10^6$ worldwide simulations of the pandemic evolution with the GLEaM model. The method shifts the burden of estimating the disease transmissibility from the prevalence data, suffering notification/surveillance biases and dependent on country specific surveillance systems, to the more accurate data of early case detection in newly affected countries.

This is achieved through the modeling of human mobility patterns on the global level obtained from high quality databases. In other words, the chronology of the infection of new countries is determined by two factors. The first is the number of cases generated by the epidemic in the originating country. The second is the mobility of people from this country to the rest of the world. The mobility data are defined from the outset with great accuracy and we can therefore find the parameters of the disease spreading as those that provide the best fit for the time of infection of new countries. This method also allows for uncovering the presence of a seasonal signature in the observed pattern, not hindered or effectively caused by notification and reporting changes in each country's influenza monitoring. The obtained values for the reproduction numbers are larger than the early estimates [5], though aligned with later works [11-13]. The simulated geographic and temporal evolution of the pandemic based on these estimates predicts an early epidemic activity peak in the Northern hemisphere as soon as early/mid October. While the simulations refer to a worst-case scenario, with no intervention implemented, the present findings pertain to the timing of the vaccination campaigns as planned by many countries. For this reason we also present an analysis of scenarios in which the systematic use of antiviral drug therapy is implemented with varying effectiveness, according to the national stockpiles, and study their effect on the epidemic timeline.

**Methods**

The GLEaM structured metapopulation model is based on a metapopulation approach [4,14-22] in which the world is divided into geographical regions defining a subpopulation network where connections among subpopulations represent the individual fluxes due to the transportation and mobility infrastructure. GLEaM integrates three different data layers [10]. The population layer is based on the high-resolution population database of the 'Gridded Population of the World' project of the SocioEconomic Data and Applications Center (SEDAC) [23] that estimates the population with a granularity given by a lattice of cells covering the whole planet at a resolution of 15 × 15 minutes of arc. The transportation mobility layer integrates air travel mobility obtained from the International Air Transport Association (IATA [24]) and Official Airline Guide (OAG [25]) databases that contain the list of worldwide airport pairs connected by direct flights and the number of available seats on any given connection [26]. The combination of the population and mobility layers allows the subdivision of the world into georeferenced census areas defined with a Voronoi tessellation procedure [27] around transportation hubs. These census areas define the subpopulations of the metapopulation modeling structure (see Figure 1). In particular, we identify 3,362 subpopulations centered around IATA airports in 220 different countries (see [10] and Additional file 1 for more details). GLEaM integrates short scale mobility between adjacent subpopulations by considering commuting patterns worldwide as obtained from the data collected and analyzed from more than 29 countries in 5 continents across the world [10]. Superimposed on these layers is the epidemic layer that defines the disease and population dynamics. The model simulates the mobility of individuals from one subpopulation to another by a stochastic procedure in which the number of passengers of each compartment traveling from a subpopulation $j$ to a subpopulation $l$ is an integer random variable defined by the actual data from the airline transportation database (see Additional file 1). Short range commuting between subpopulations is modeled with a time scale separation approach that defines the effective force of infections in connected subpopulations [10,28,29]. The infection dynamics takes place within each subpopulation and assumes the classic influenza-like illness compartmentalization in which each individual is classified by a discrete state such as susceptible, latent, infectious symptomatic, infectious non-symptomatic or permanently recovered/removed [9,30]. The model therefore assumes that the latent period is equivalent to the incubation period and that no secondary transmissions occur during the incubation period (see Figure 1 for a detailed description of the compartmentalization). All transitions are modeled through binomial and multinomial processes to preserve the discrete and stochastic nature of the processes (see Additional file 1 for the full description). Asymptomatic individuals are considered as a fraction $p_a$ = 33% of the infectious individuals [31] generated in the model and assumed to infect with a relative infectiousness of $r_\beta$ = 50% [5,30,32]. Change in traveling behavior after the onset of symptoms is modeled with the probability 1 - $p_t$, set to 50%, that individuals would stop traveling when ill [30]. The spreading rate of the disease is ultimately governed by the basic reproduction number $R_0$. Once the disease parameters and initial conditions based on available data are defined, GLEaM allows the generation of stochastic realizations of the worldwide unfolding of the epidemic, with mobility processes entirely based on real data. The model generates *in silico* epidemics for which we can gather information such as prevalence, morbidity, number of secondary cases, number of imported cases and other quantities for each subpopulation and with a time resolution of 1 day. While global models are generally used to produce scenarios in which the basic disease parameters are defined from the outset, here we use the model to provide a maximum likelihood estimate of the transmission potential by finding the set of disease parameters that best fit the data on the arrival time of cases in different countries worldwide. It is important to stress that the model is not an agent-based model and does not include additional structure within a subpopulation, therefore it cannot provide detailed information at the level of households or workplaces. The projections for the winter season in the northern hemisphere are also assuming that there will be no mutation of the virus with respect to the spring/summer of 2009. Furthermore, while at the moment the novel H1N1 influenza is accounting for

75% of the influenza cases worldwide, the model does not consider the cocirculation of different influenza strains and cannot provide information on cocirculation data.

The initial conditions of the epidemic are defined by setting the onset of the outbreak near La Gloria in Mexico on 18 February 2009, as reported by official sources [33] and analogously to other works [5]. We tested different localizations of the first cases in census areas close to La Gloria without observing relevant variations with respect to the observed results. We also performed sensitivity analysis on the starting date by selecting a seeding date anticipated or delayed by 1 week with respect to the date available in official reports [33]. The arrival time of infected individuals in the countries seeded by Mexico is clearly a combination of the number of cases present in the originating country (Mexico) and the mobility network, both within Mexico and connecting Mexico with countries abroad. For this reason we integrated into our model the data on Mexico-US border commuting (see Figure 2a), which could be relevant in defining the importation of cases in the US, along with Mexican internal commuting patterns (see Figure 1) that are responsible for the diffusion of the disease from rural areas as La Gloria to transportation hubs such as Mexico City. In addition, we used a time-dependent modification of the reproductive number in Mexico as in [6] to model the control measures implemented in the country starting 24 April and ending 10 May, as those might affect the spread to other countries.

In order to ascertain the effect of seasonality on the observed pattern, we explored different seasonality schemes. The seasonality is modeled by a standard forcing that rescales the value of the basic reproductive number into a seasonally rescaled reproductive number, $R(t)$, depending on time. The seasonal rescaling is time and location dependent by means of a scaling multiplicative factor generated by a sinusoidal function with a total period of 12 months oscillating in the range $\alpha_{min}$ to $\alpha_{max}$, with $\alpha_{max} = 1.1$ (sensitivity analysis in the range 1.0 to 1.1) and $\alpha_{min}$ a free parameter to be estimated [17]. The rescaling function is in opposition in the Northern and Southern hemispheres (see Additional file 1 for details). No rescaling is assumed in the Tropics. The value of $R_0$ reported in the Tables and the definition of the baseline is the reference value in the Tropics. In each subpopulation the $R(t)$ relative to the corresponding geographical location and time of the year is used in the simulations.

The seasonal transmission potential of the H1N1 strain is assessed in a two-step process that first estimates the reproductive number in the Tropics region, where seasonality is assumed not to occur, by focusing on the early international seeding by Mexico, and then estimates the degree of seasonal dumping factor by examining a longer time period of international spread to allow for seasonal changes. The estimation of the reproductive number is performed through a maximum likelihood analysis of the model fitting the data of the early chronology of the H1N1 epidemic. Given a set of values of the disease parameters, we produced $2 \times 10^3$ stochastic realizations of the pandemic evolution worldwide for each $R_0$ value. Our model explicitly takes into account the class of symptomatic and asymptomatic individuals (see Figure 1) and allows the tracking of the importation of each symptomatic individual and of the onset of symptoms of exposed individuals transitioning to the symptomatic class, as observables of the simulations. This allows us to obtain numerically with a Monte Carlo procedure the probability distribution $P_i(t_i)$ of the importation of the first infected individual or the first occurrence of the onset of symptoms for an individual in each country i at time $t_i$. Asymptomatic individuals do not contribute to the definition of $t_i$. With the aim of working with conditional independent variables we restrict the likelihood analysis to 12 countries seeded from Mexico (see Figure 2b) and for which it is possible to know with good confidence the onset of symptoms and/or the arrival date of the first detected case (see Tables and data sources in Additional file 1). This allows us to define a likelihood function $L = \Pi_i P_i(t_i^*)$, where $t_i^*$ is the empirical arrival time from the H1N1 chronological history in each of the selected countries. This methodology assumes the prompt detection of symptomatic cases at the very beginning of the outbreak in a given country, and for this reason we have also provided a sensitivity analysis accounting for a late/missed detection of symptomatic individuals as reported in the next section. The transmission potential is estimated as the value of $R_0$ that maximizes the likelihood function $L$, for a given set of values of the disease parameters. In Table 1 we report the reference values assumed for some of the model parameters and the range explored with the sensitivity analysis. So far there are no precise clinical estimates of the basic model parameters $\varepsilon$ and $\mu$ defining the inverse average exposed and infectious time durations [34-36]. The generation interval $G_t$ [37,38] used in the literature is based on the early estimate of [5] and values obtained for previous pandemic and seasonal influenza [4,30-32,39,40], with most studies focusing on values ranging from 2 to 4 days [5,11-13]. We have therefore assumed a short exposed period value $\varepsilon^{-1} = 1.1$ as indicated by early estimates [5] and compatible with recent studies on seasonal influenza [31,41] and performed a sensitivity analysis for values as large as $\varepsilon^{-1} = 2.5$. The maximum likelihood procedure is performed by systematically exploring different values of the generation time aimed at providing a best estimate and confidence interval for $G_t$, along with the estimation of the maximum likelihood value of $R_0$.

The major problem in the case of projections on an extended time horizon is the seasonality effect that in the long run is crucial in determining the peak of the epidemic. In order to quantify the degree of seasonality observed in the current epidemic, we estimate the minimum seasonality scaling factor $\alpha_{min}$ of the sinusoidal forcing by extending the chronology under study and analyzing the whole data set composed of the arrival dates of the first infected case in the 93 countries affected by the outbreak as of 18 June. We studied the correlation between the simulated arrival time by country and its corresponding empirical value, by measuring the regression coefficient between the two datasets. Given the extended time frame under observation, the arrival times considered in this case are expected to provide a signature of the presence of seasonality. They included the seeding of new countries from outbreaks taking place in regions where seasonal effects might occur, as for example in the US or in the UK. For the simulated arrival times we have considered the median and 95% confidence interval (CI) emerging from the $2 \times 10^3$ stochastic runs. The regression coefficient is found to be sensitive to variations in the seasonality scaling factor, allowing discrimination of the $\alpha_{min}$ value that best fits the real epidemic. A detailed presentation of this analysis is reported in Additional file 1. The full exploration of the phase space of epidemic parameters and seasonality scenarios reported in Additional file 1 required data from $10^6$ simulations; the equivalent of 2 million minutes of PowerPC 970 2,500 MHz CPU time.

**Results and Discussion**

Table 1 reports the results of the maximum likelihood procedure and of the correlation analysis on the arrival times for the estimation of $\alpha_{min}$. In the following we consider as the baseline case the set of parameters defined by the best estimates: $G_t$ = 3.6 days, $\mu^{-1}$ = 2.5 days, $R_0$ = 1.75.

The best estimates for $G_t$ and $R_0$ are higher than those obtained in early findings but close to subsequent analysis on local outbreaks [11-13]. The $R_0$ we report is the reference value for Mexico and the tropical region, whereas in each country we have to consider the $R(t)$ due to the seasonality rescaling depending on the time of the year, as shown in Table 2. This might explain the lower values found in some early analysis in the US. The transmission potential emerging from our analysis is close to estimates for previous pandemics [14,42]. In Additional file 1 we provide supplementary tables for the full sensitivity analysis concerning the assumptions used in the model. Results show that larger values of the generation interval provide increasing estimates for $R_0$. Fixing the latency period to $\varepsilon^{-1}$ = 1.1 days and varying the mean infectious period in the plausible range 1.1 to 4.0 days yields corresponding maximum likelihood estimates for $R_0$ in the range 1.4 to 2.1. Variations in the latency period from $\varepsilon^{-1}$ = 1.1 to $\varepsilon^{-1}$ = 2.5 provide corresponding best estimates for $R_0$ in the range 1.9 to 2.3, if we assume an infectious period of 3 days. We tested variations of the compartmental model parameters $p_a$, and $p_t$ up to 20% and explored the range $r_\beta$ = 20% to 80%, and sensitivity on the value of the maximum seasonality scaling factor $\alpha_{max}$ in the range 1.0 to 1.1. The obtained estimates lie within the confidence intervals of the best estimate values.

The empirical arrival time data used for the likelihood analysis are necessarily an overestimation of the actual date of the importation of cases as cases could go undetected. If we assume a shift of 7 days earlier for all arrival times available from official reports, the resulting maximum likelihood is increasing the best estimate for $R_0$ to 1.87 (95% CI 1.73 to 2.01), as expected since earlier case importation necessitates a larger growth rate of the epidemic. The official timeline used here therefore provides, all other parameters being equal, a lower estimate of the transmission potential. We have also explored the use of a subset of the 12 countries, always generating results within the confidence interval of the best estimate.

The best estimates reported in Table 1 do not show an observable dependence on the assumption about the seasonality scenario (as reported in Additional file 1). The analysis is restricted to the first countries seeded from Mexico to preserve the conditional independence of the variables and it is natural to see the lack of any seasonal signature since these countries receive the disease from a single country, mostly found in the tropical region where no seasonal effects are expected.

In order to find the minimum seasonality scaling factor $\alpha_{min}$ that best fits the empirical data, we performed a statistical correlation analysis of the arrival time of the infection in the 93 countries infected as of 18 June, as detailed in the Methods section and Additional file 1. By considering a larger number of countries and a longer period for the unfolding of the epidemic worldwide as seasons change, the correlation analysis for the baseline scenario provides clear statistical indications for a minimum rescaling factor in the interval $0.6 < \alpha_{min} < 0.7$. In the full range of epidemic parameters explored, the correlation analysis yields values for $\alpha_{min}$ in the range 0.4 to 0.9. This evidence for a mild seasonality rescaling is consistent with the activity observed in the months of June and July in Europe and the US where the epidemic progression has not stopped and the number of cases keeps increasing considerably (see also Table 2 for the corresponding values of $R(t)$ in those regions during summer months).

This analysis allows us to provide a comparison with the epidemic activity observed so far, and most importantly an early assessment of the future unfolding of the epidemics. For each set of parameters the model generates quantities of interest such as the profile of the epidemic behavior in each subpopulation or the number of imported cases. Each simulation generates a stochastic realization of the process and the curves are the statistical aggregate of at least $2 \times 10^3$ realizations. In the following we report the median profiles and where indicated the 95% CI. For the sake of clarity data are aggregated at the level of country or geographical region. Additional file 1 reports a detailed comparison of the simulated number of cases in Australia, US, UK with the reported cases from official sources in the period May to July. Results are in good agreement with the reported temporal evolution of the epidemic and highlight a progressive decrease of the monitoring activity caused by the increasing number of cases, as expected [7]. The same information is also available for each single subpopulation defined in the model. We have therefore tested the model results in four territories of Australia. Interestingly, the model is able to recover the different timing observed in the four territories. A detailed discussion of this comparison is reported in Additional file 1.

In Figure 2c-d we report the predicted baseline case profiles for countries in the Southern hemisphere. It is possible to observe in the figure that in this case, the effect of seasonality is not discriminating between different waves, as the short time interval from the start of the outbreak to the winter season in the Southern hemisphere does not allow a large variation in the rescaling of the transmissibility during these months. Therefore we predict a first wave that occurs between August and September in phase with the seasonal influenza pattern, and independently of the seasonality parameter $\alpha_{min}$. The situation is expected to be different in the Northern hemisphere where different seasonality parameters might progressively shift the peak of the epidemic activity in the winter months. Figure 2e reports the predicted daily incidence profiles for the Northern hemisphere and the 95% CI for the activity peaks of the pandemic with the best-fit seasonality scenario (that is, the range 0.6 <

$\alpha_{min} < 0.7$). Table 3 reports the same information for different continental areas. The general evidence clearly points to the occurrence of an autumn/winter wave in the Northern hemisphere strikingly earlier than expected, with peak times ranging from early October to the middle of November. The peak estimate for each geographical area is obtained from the epidemic profile summing up all subpopulations belonging to the region. The activity peak estimate for each single country can be noticeably different from the overall estimate of the corresponding geographical region as more populated areas may dominate the estimate for a given area. For instance Chile has a pandemic activity peak in the interval 1 July – 6 August, one month earlier than the average peak estimate for the Lower South America geographical area it belongs to. It is extremely important to remark that in the whole phase space of parameters explored the peak time for the epidemic activity in the Northern hemisphere lies in the range late September to late November, thus suggesting that the early seasonal peak is a genuine feature induced by the epidemic data available so far.

In Table 4 we report the new number of cases at the activity peak and the epidemic size as of 15 October for a selected number of countries. As shown by the results in the table, the implementation of a massive vaccination campaign starting in October or November, with no additional mitigation implemented, would be too late with respect to the epidemic evolution, and could therefore be expected to be rather ineffective in reducing transmission. This makes a strong case for prioritized vaccination programs focusing on high-risk groups and healthcare and social infrastructures workers. In order to assess the amount of pressure on the healthcare infrastructure, in Table 5 we provide the expected number of hospitalizations at the epidemic peak according to different hospitalization rate estimates. The assessment of the hospitalization rate is very difficult as it depends on the ratio between the number of hospitalizations and the actual number of infected people. As discussed previously, the number of confirmed cases released by official agencies is always a crude underestimate of the actual number of infected people. We consider three different methods along the lines of those developed for the analysis of fatalities due to the new virus [43]. The first assumes the average value of hospitalization observed during the regular seasonal influenza season. The second is a multiplier method in which the hospitalization rate is obtained as the ratio between the WHO number of confirmed hospitalizations and the cases confirmed by the WHO multiplied by a factor 10 to 30 to account for underreporting. The third method is given by the ratio of the total number of confirmed hospitalizations and the total number of confirmed cases. This number is surely a gross overestimation of the hospitalization rate [43,44]. It has to be noted that hospitalizations are often related to existing health conditions, age and other risk factors. This implies that hospitalizations will likely not affect the population homogenously, a factor that we cannot consider in our model.

The number of hospitalized at peak times in the selected countries range between 2 and 40 per 100,000 persons, for a hospitalization rate typical of seasonal influenza and for an assumed 1% rate, respectively, yielding a quantitative indication of the potential burden that the health care systems will likely face at the peak of the epidemic activity in the next few months. It is worth noting that the present analysis considers a worst-case scenario in which no effective containment measures are introduced. This is surely not the case in that pandemic plans and mitigation strategies are considered at the national and international level. Guidelines aimed at increasing social distancing and the isolation of cases will be crucial in trying to mitigate and delay the spread in the community, thus reducing the overwhelming requests on the hospital systems. Most importantly, the mass vaccination of a large fraction of the population would strongly alter the presented picture. By contrast, any mass vaccination campaign is unlikely to start before the middle of October [45,46]. The potential for an early activity peak of the pandemic in October/November puts at risk the effectiveness of any mass vaccination program that might take place too late with respect to the pandemic wave in the Northern hemisphere. In this case it is natural to imagine the use of other mitigation strategies aimed at delaying the activity peak so that the maximum benefit can be gained with the vaccination program. As an example, we studied the implementation of systematic antiviral (AV) treatment and its effect in delaying the activity peak [19,30,32,39,47-50]. The resulting effects are clearly country specific in that each country will experience a different timing for the epidemic peak (with a local transmissibility increasing in value as we approach the winter months) and will count on antiviral stockpiles of different sizes. Here we consider the implementation of the AV treatment in all countries in the world that have drugs stockpiles available (source data from [51,52] and national agencies), until the exhaustion of their stockpiles [4]. We have modeled this mitigation policy with a conservative therapeutic successful use of drugs for 30% of symptomatic infectious individuals. The efficacy of the AV is accounted in the model by a 60% reduction in the transmissibility of the disease of an infected person under AV treatment when AV drugs are administered in a timely fashion [30,32]. We assume that the drugs are administered within 1 day of the onset of symptoms. We also consider that the AV treatment reduces the infectious period by 1 day [30,32]. In Figure 3 we show the delay obtained with the implementation of the AV treatment protocol in a subset of countries with available stockpiles. As an example, we also show the incidence profiles for the cases of Spain and Germany, where it is possible to achieve a delay of about 4 weeks with the use of 5 million and 10 million courses of AV, respectively. The results of this mitigation might be extremely valuable in providing the necessary time for the implementation of the mass vaccination program.

**Conclusions**

We have defined a Monte Carlo likelihood analysis for the assessment of the seasonal transmission potential of the new A(H1N1) influenza based on the analysis of the chronology of case detection in affected countries at the early stage of the epidemic. This method allows the use of data coming from the border controls and the enhanced surveillance aimed at detecting the first cases reaching uninfected countries. This data is, in principle, more reliable than the raw count of cases provided by countries during the evolution of the epidemic. The procedure provides the necessary input to the large-scale computational model for the analysis of the unfolding of the pandemic in the future months. The analysis shows the potential for an early activity peak that strongly emphasizes the need for detailed planning for additional intervention measures, such as social distancing and antiviral drugs use, to delay the epidemic activity peak and thus increase the effectiveness of the subsequent vaccination effort.

## Competing interests

The authors declare they have no competing interests.

## Authors' contributions

DB, HH, BG, PB and CP contributed to conceiving and designing the study, performed numerical simulations and statistical analysis, contributed to the data integration and helped to draft the manuscript. JJR contributed to conceiving and designing the study, data tracking and integration, statistical analysis and helped draft the manuscript. NP and MT contributed to data tracking and integration, statistical analysis and helped draft the manuscript. DP contributed to data integration and management and helped draft the manuscript. WVdB contributed to visualization and data management. AV and VC conceived, designed and coordinated the study, contributed to the analysis and methods development and drafted the manuscript. All authors read and approved the final manuscript.

## Acknowledgements

The authors thank IATA and OAG for providing their databases. The authors are grateful to the Staff of the Big Red Computer and the Computational Facilities at Indiana University. The authors would like to thank Ciro Cattuto for his support with computational infrastructure at ISI Foundation. The authors are partially supported by the NIH, the NSF, the Lilly Endowment Foundation, DTRA, the ERC project EpiFor and the FET projects Epiwork and Dynanets.

**Figure legends**

**Figure 1**

**Schematic illustration of the GLobal Epidemic and Mobility (GLEaM) model.** Top: census and mobility layers that define the subpopulations and the various types of mobility among those (commuting patterns and air travel flows). The same resolution is used worldwide. Bottom: compartmental structure in each subpopulation. A susceptible individual in contact with a symptomatic or asymptomatic infectious person contracts the infection at rate $\beta$ or $r_\beta\beta$ [30,32], respectively, and enters the latent compartment where he is infected but not yet infectious. At the end of the latency period, each latent individual becomes infectious, entering the symptomatic compartments with probability $1 - p_a$ or becoming asymptomatic with probability $p_a$ [30,32]. The symptomatic cases are further divided between those who are allowed to travel (with probability $p_t$) and those who would stop traveling when ill (with probability $1 - p_t$) [30]. Infectious individuals recover permanently with rate $\mu$. All transition processes are modeled through multinomial processes.

**Figure 2.**

**Illustration of the model's initialization and the results for the activity peaks in three geographical areas. (a)** Intensity of the commuting between US and Mexico at the border of the two countries. **(b)** The 12 countries infected from Mexico used in the Monte Carlo likelihood analysis. The color scale of the arrows from red to yellow indicates the time ordering of the epidemic invasion. Panels **(c)**, **(d)** and **(e)** show the daily incidence in Lower South America, South Pacific and North America/Western Europe, respectively. The shaded area indicates the 95% confidence interval (CI) of the peak time in the corresponding geographical region. The median incidence profiles of selected countries are shown for the two values defining the best-fit seasonality scaling factor interval.

**Figure 3**

**Delay effect induced by the use of antiviral drugs for treatment with 30% case detection and drug administration. (a)** Peak times of the epidemic activity in the worst-case scenario (black) and in the scenario where antiviral treatment is considered (red), for a set of countries in the Northern hemisphere. The intervals correspond to the 95% confidence interval (CI) of the peak time for the two values defining the best-fit seasonality scaling factor interval. **(b,c)** Incidence profiles for Spain and Germany in the worst-case scenario (black) and in the scenario where antiviral treatment is considered (red). Results are shown for $\alpha_{min} = 0.6$ only, for the sake of visualization. A delay of about 4 weeks results from the implemented mitigation.

**Table 1. Best Estimates of the epidemiological parameters**

| Parameter | Best Estimate | Interval estimate[a] | Description |
|---|---|---|---|
| $R_0$ | 1.75 | 1.64 to 1.88 | Basic reproduction number |
| $G_t$ | 3.6 | 2.2 to 5.1 | Mean generation time (days) |

| | Assumed value at best estimate | Sensitivity analysis range | |
|---|---|---|---|
| $\mu^{-1}$ | 2.5 | 1.1 to 4.0 | Mean infectious period (days) |
| $\alpha_{min}$ | 0.65 | 0.6 to 0.7 | Minimal seasonality rescaling |

| | Assumed values: | | |
|---|---|---|---|
| | Assumed value at best estimate | Sensitivity analysis range | |
| $\varepsilon^{-1}$ | 1.1 | 1.1 to 2.5 | Mean exposed period (days) |
| $\alpha_{max}$ | 1.1 | 1.0 to 1.1 | Maximum seasonality rescaling |

Estimates from the Monte Carlo likelihood analyses for various values of the parameter space explored. In Additional file 1 we report the complete tables corresponding to the sensitivity analysis. (a) For $R_0$, we report the 95% Confidence Interval. $G_t$, $\mu^{-1}$ intervals are defined by the range of plausible constrained values sampled in the Monte Carlo approach that satisfy a likelihood ratio test at the 5% level. The $\alpha_{min}$ interval is the best-fit range within the minimal resolution allowed by the Montecarlo sampling.

**Table 2. Seasonality time-dependent reproduction number in the Northern hemisphere**

| Month | $R$(t) in Northern hemisphere |
|---|---|
| May | 1.19 to 1.49 |
| June | 1.07 to 1.33 |
| July | 1.05 to 1.24 |
| August | 1.07 to 1.33 |
| September | 1.19 to 1.49 |

The values of $R$(t) for the Northern hemisphere correspond to the rescaling of the maximum likelihood value of $R_0$ in Mexico and in the Tropical regions ($R_0$ = 1.75) and the best values for the seasonality rescaling factor, $0.6 < \alpha_{min} < 0.7$. The parameter $\alpha_{min}$ indicates the minimum value of the seasonal rescaling of $R_0$ induced by the sinusoidal forcing in the Northern hemisphere [17].

**Table 3. Peak times**

| Region | Estimated activity peak time |
| --- | --- |
| North America | 25 September to 9 November |
| Western Europe | 14 October to 21 November |
| Lower South America | 30 July to 6 September |
| South Pacific | 28 July to 17 September |

The table reports the 95% confidence interval (CI) for the pandemic activity peak time for geographical areas in the Northern and Southern hemispheres estimated for the best-fit seasonality scaling interval, $0.6 < \alpha_{min} < 0.7$, and for the maximum likelihood value of $R_0$ found for the baseline parameters, $R_0 = 1.75$. The confidence interval is obtained from the set of numerical observations of the peak time in a given region obtained from the 2,000 stochastic runs of the model. In Additional file 1 we report the results for the full sensitivity analysis. In all cases we obtain activity peak time intervals close to those reported for the baseline scenario. Peak time estimates in this table are obtained from the epidemic profile of the entire geographical region. Single country belonging to each region could have different peak time estimates (see text).

**Table 4: Daily new number of cases and epidemic sizes in several countries**

| Country | Peak time | New daily cases at the peak time (thousands) | New daily cases at the peak time (% of population) | Epidemic size at 15 October (% of population) | |
|---|---|---|---|---|---|
| | | | | $\alpha_{min}$ 0.6 | $\alpha_{min}$ 0.7 |
| United States | 24 September to 9 November | 2,983 to 3,302 | 1.06 to 1.17 | 4.99 to 7.38 | 23.76 to 29.96 |
| Canada | 4 October to 14 November | 331 to 373 | 1.04 to 1.17 | 2.28 to 4.56 | 16.90 to 27.41 |
| United Kingdom | 9 October to 18 November | 723 to 813 | 1.21 to 1.36 | 1.77 to 4.45 | 11.11 to 27.29 |
| France | 12 October to 21 November | 725 to 792 | 1.26 to 1.38 | 1.83 to 3.87 | 10.86 to 26.40 |
| Germany | 11 October to 20 November | 1,162 to 1,291 | 1.43 to 1.59 | 1.02 to 2.41 | 8.57 to 26.25 |
| Italy | 17 October to 23 November | 793 to 867 | 1.39 to 1.52 | 0.93 to 2.20 | 6.71 to 22.13 |
| Spain | 8 October to 19 November | 492 to 536 | 1.23 to 1.34 | 2.39 to 3.70 | 13.26 to 27.95 |
| China | 8 November to 11 December | 14,077 to 16,207 | 1.16 to 1.34 | 0.65 to 5.34 | 1.51 to 9.49 |
| Japan | 13 October to 16 November | 1,539 to 1,822 | 1.21 to 1.43 | 1.47 to 4.86 | 5.84 to 24.65 |

Peak times of the epidemic activity, daily new number of cases predicted at peak time and % of the population, and epidemic size on 15 October are shown. Intervals refer to the 95% confidence interval (CI). After 1 year from the start of the epidemic the percentage of total population infected is close to 45% with small differences of the order of 5% across different countries.

**Table 5. Number of hospitalizations per 100,000 persons at the activity peak in several countries**

| | HR based on seasonal influenza, 0.08% | HR based on multiplier method | | HR based on WHO confirmed cases, 10% |
| --- | --- | --- | --- | --- |
| | | 0.3% | 1% | |
| USA | 2.21 | 8.28 | 27.58 | 275.84 |
| Canada | 2.18 | 8.17 | 27.22 | 272.23 |
| UK | 2.52 | 9.45 | 31.52 | 315.15 |
| France | 2.61 | 9.79 | 32.64 | 326.40 |
| Germany | 2.98 | 11.17 | 37.22 | 372.18 |
| Italy | 2.87 | 10.76 | 35.87 | 358.67 |
| Spain | 2.54 | 9.54 | 31.81 | 318.12 |
| China | 2.48 | 9.32 | 31.05 | 310.50 |
| Japan | 2.59 | 9.70 | 32.32 | 323.19 |

The estimates are obtained by considering three methods. The first assumes the average hospitalization rate (HR) observed during the seasonal influenza season. The second is a simple multiplier method in which the HR is obtained as the ratio between the World Health organization (WHO) number of confirmed hospitalizations and the cases confirmed by the WHO multiplied by a factor 10 to 30 to account for underreporting. The third method is simply the ratio of the total number of confirmed hospitalizations and the total number of confirmed cases.

**Additional files**

**Additional file 1**

File format: PDF

Title: Additional information.

Description: The file provides details on the model and all the statistical and sensitivity analysis carried out in the preparation of this work. The file also contains references to all data sources used in the preparation of this work.

## population layer

## mobility layers

*geographic scale*

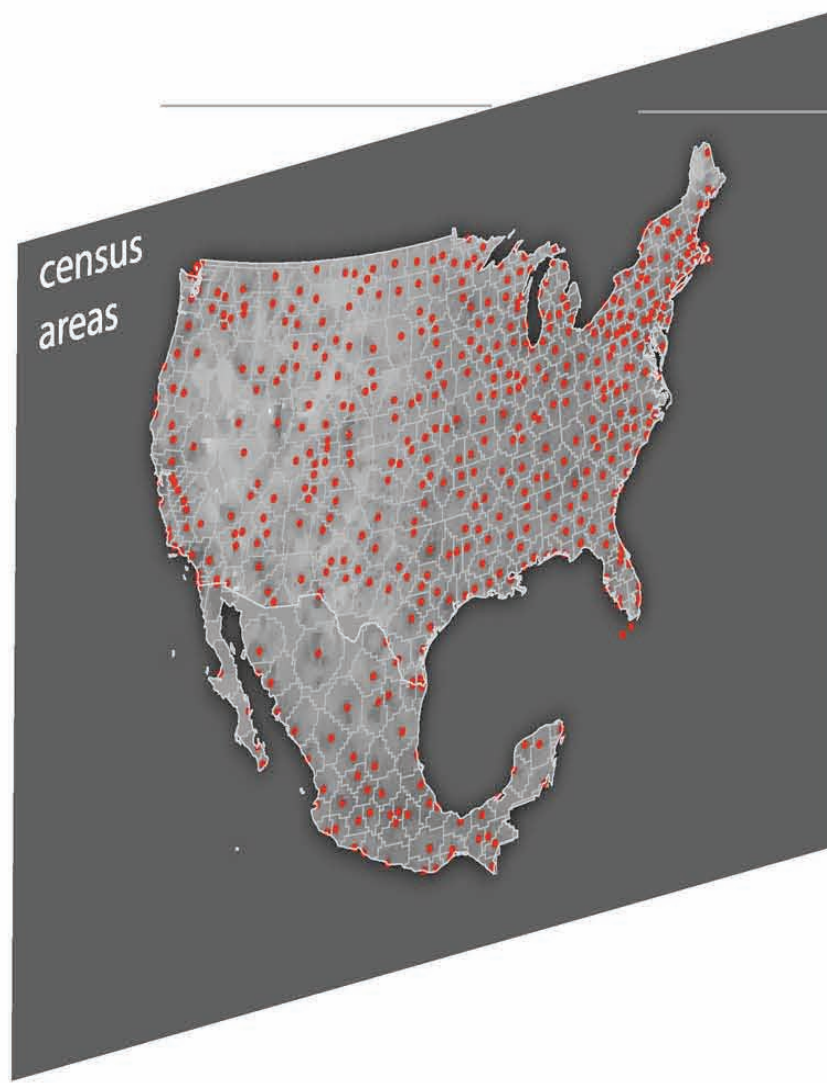

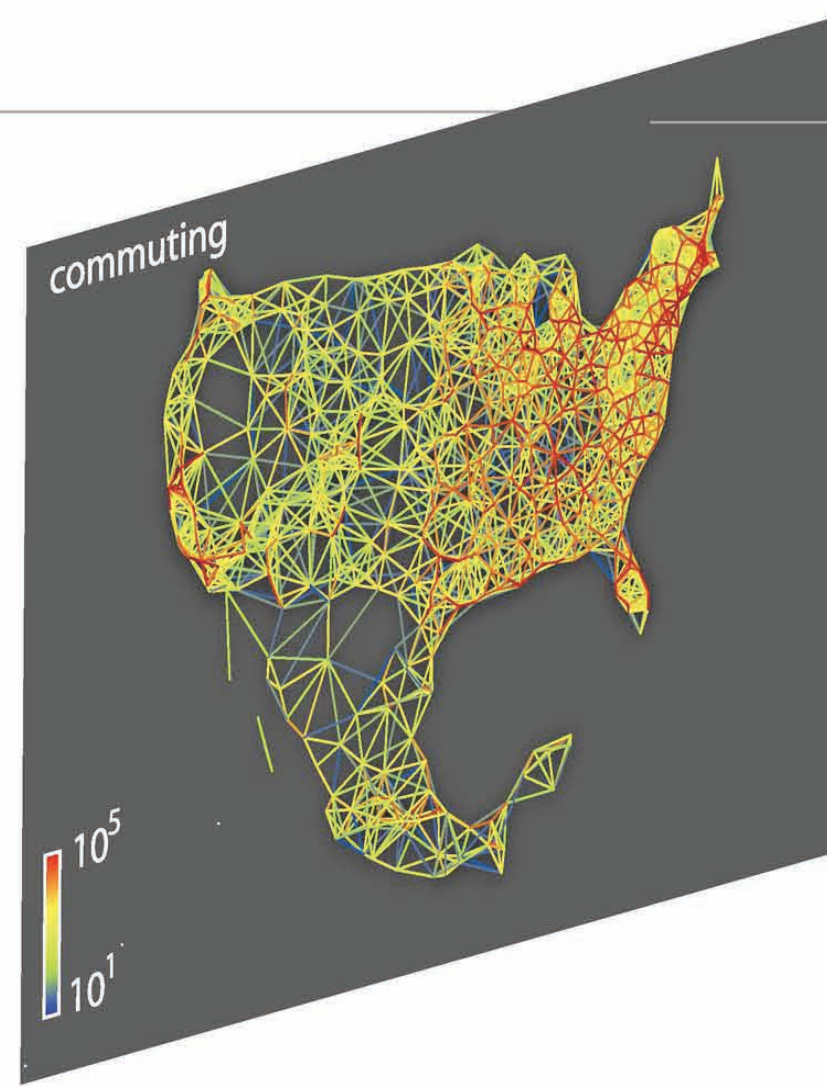

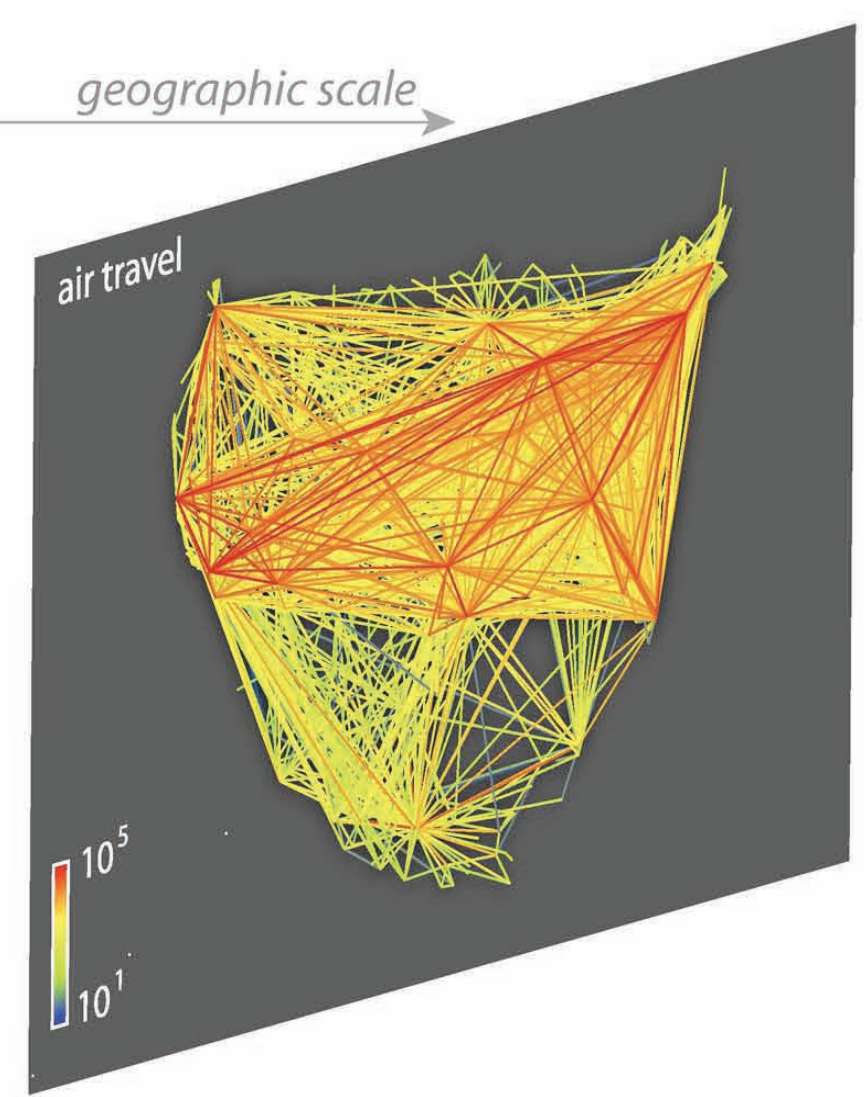

## epidemic layer

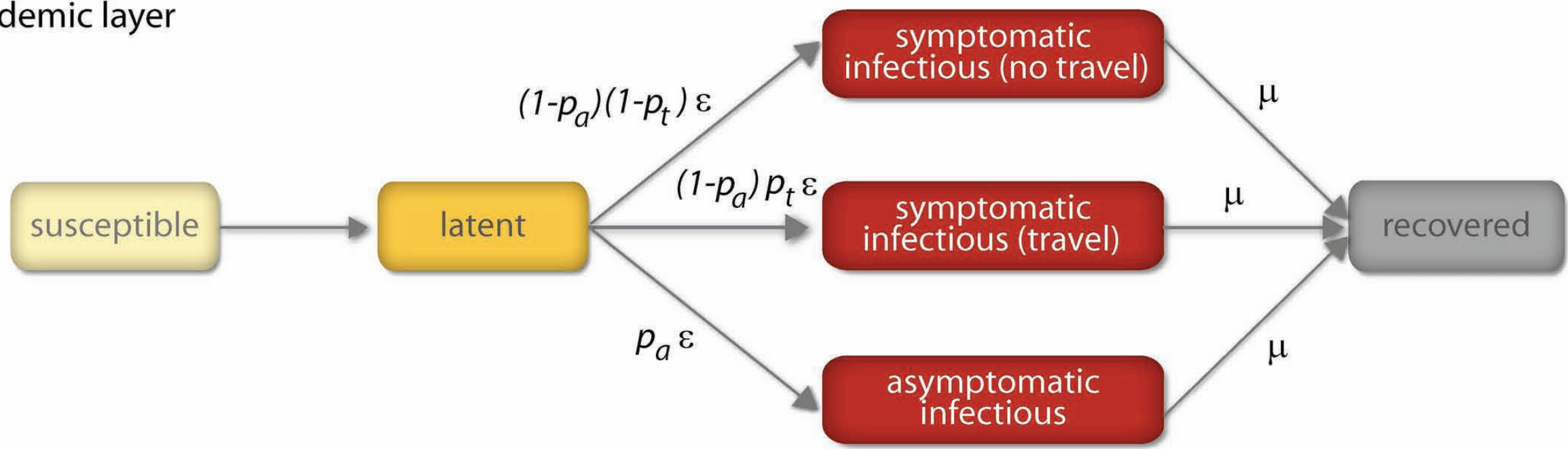

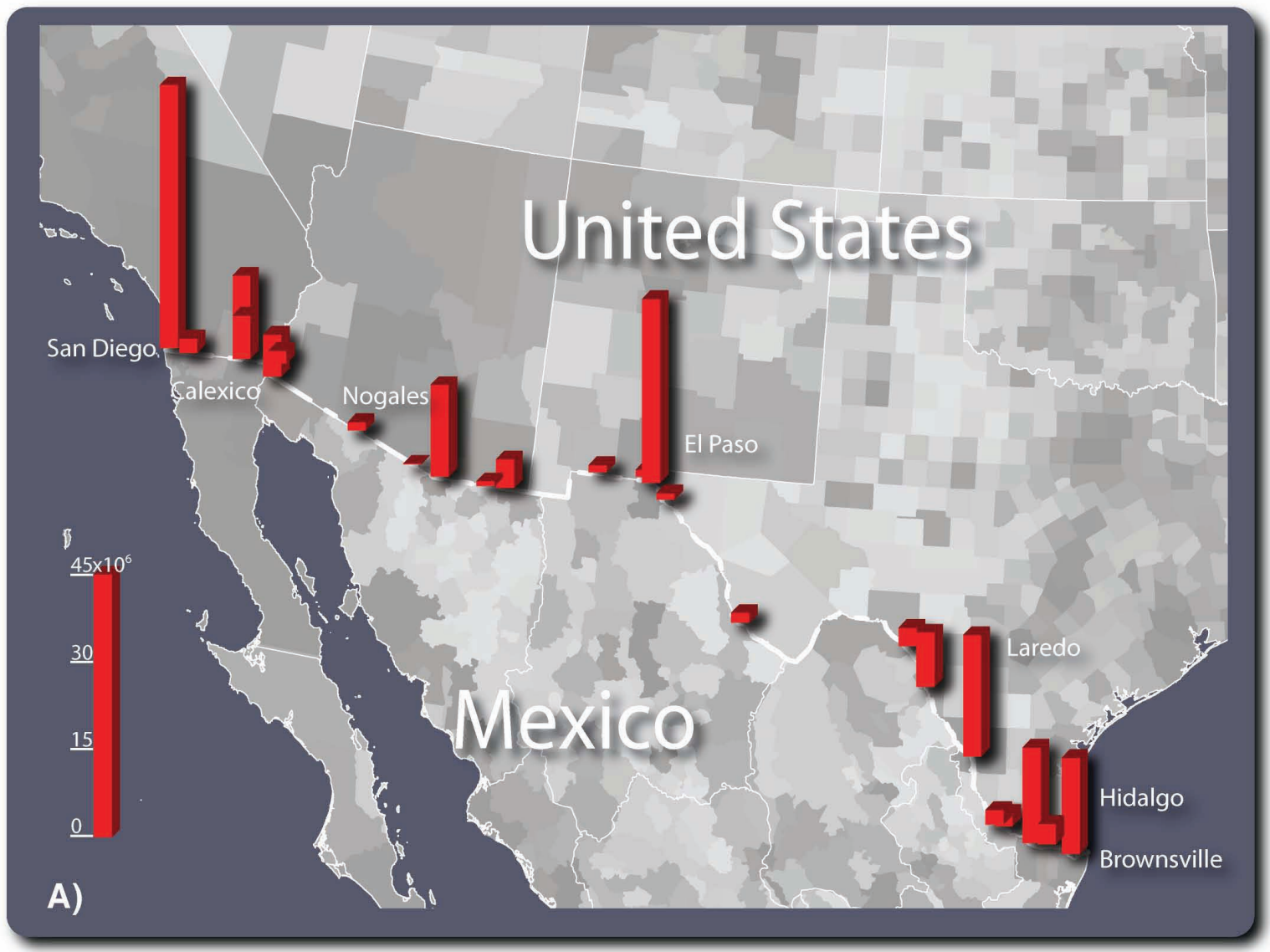

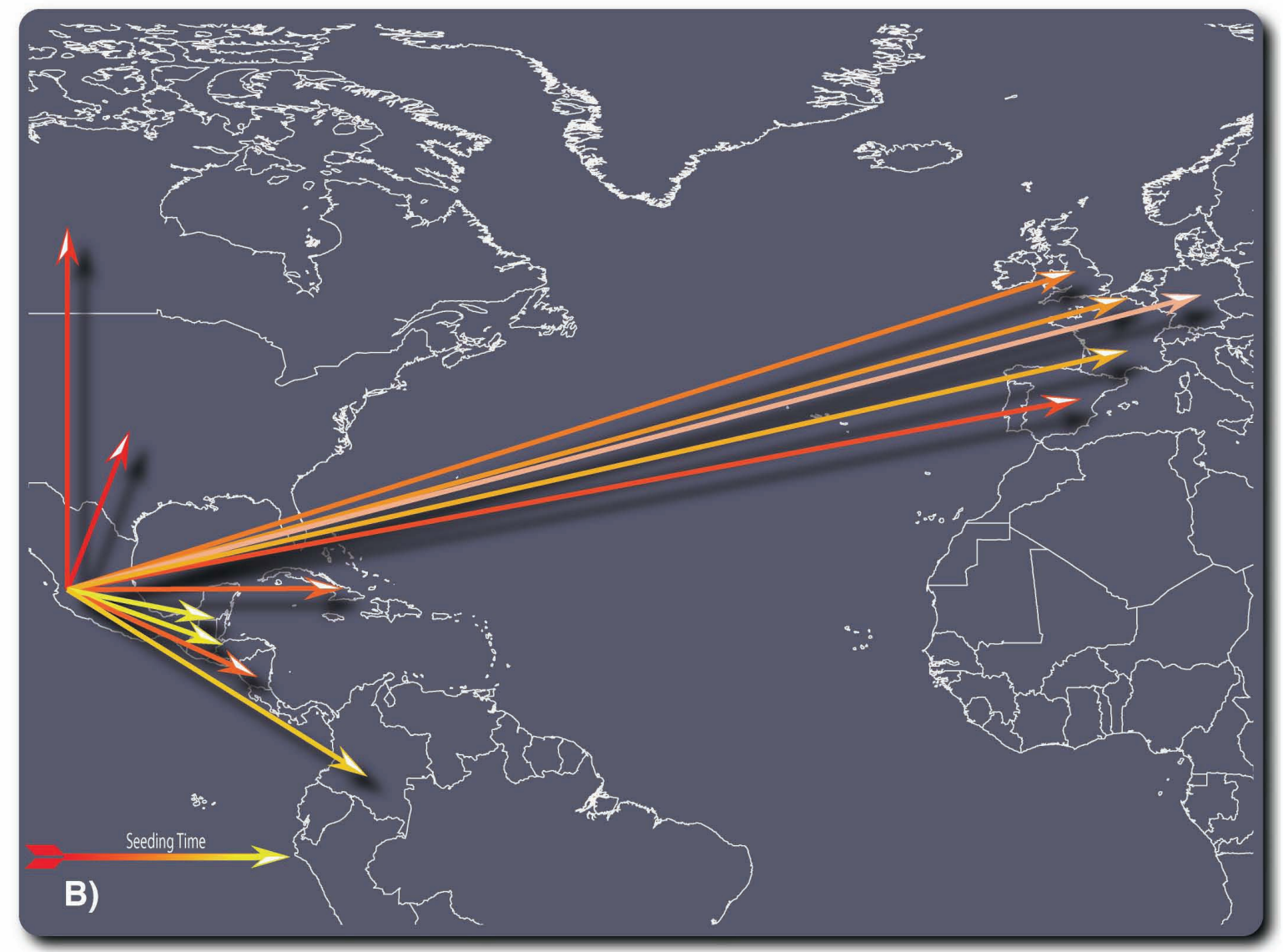

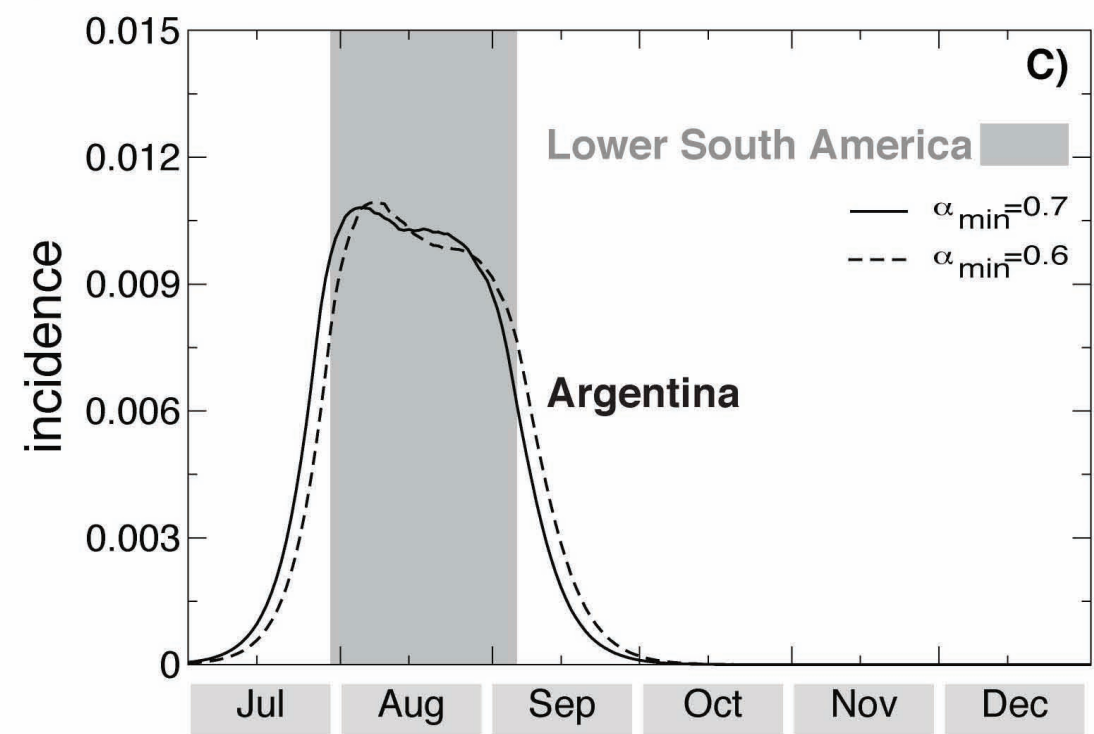

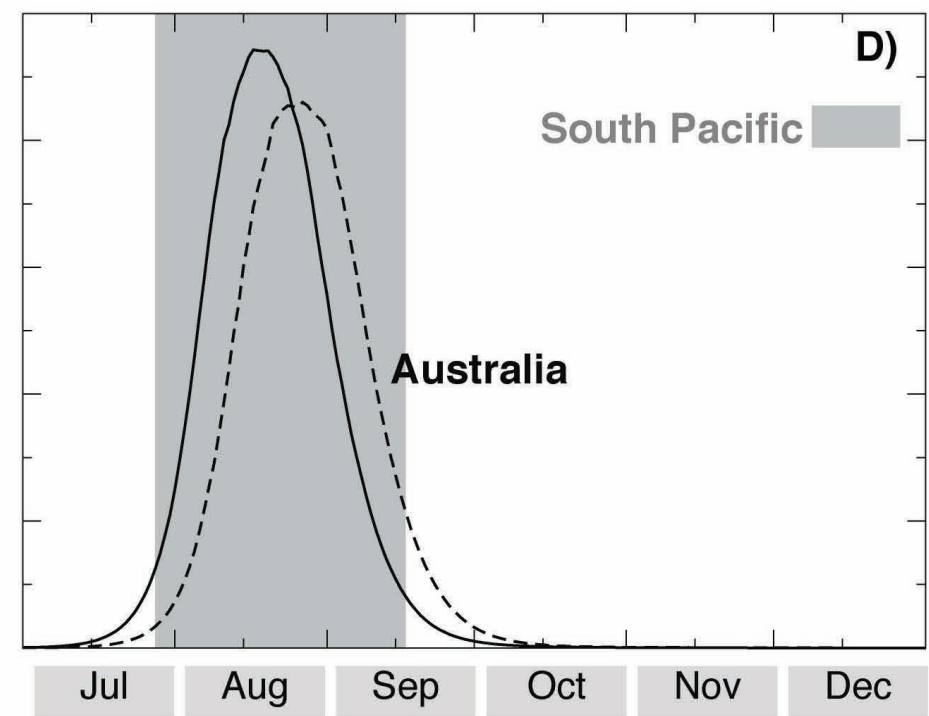

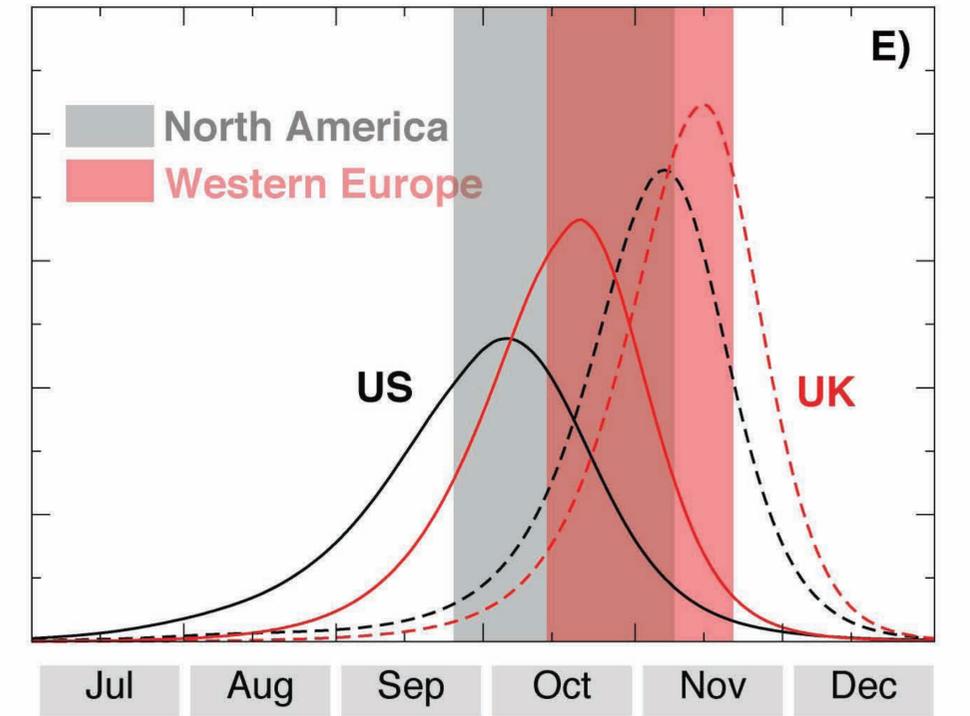

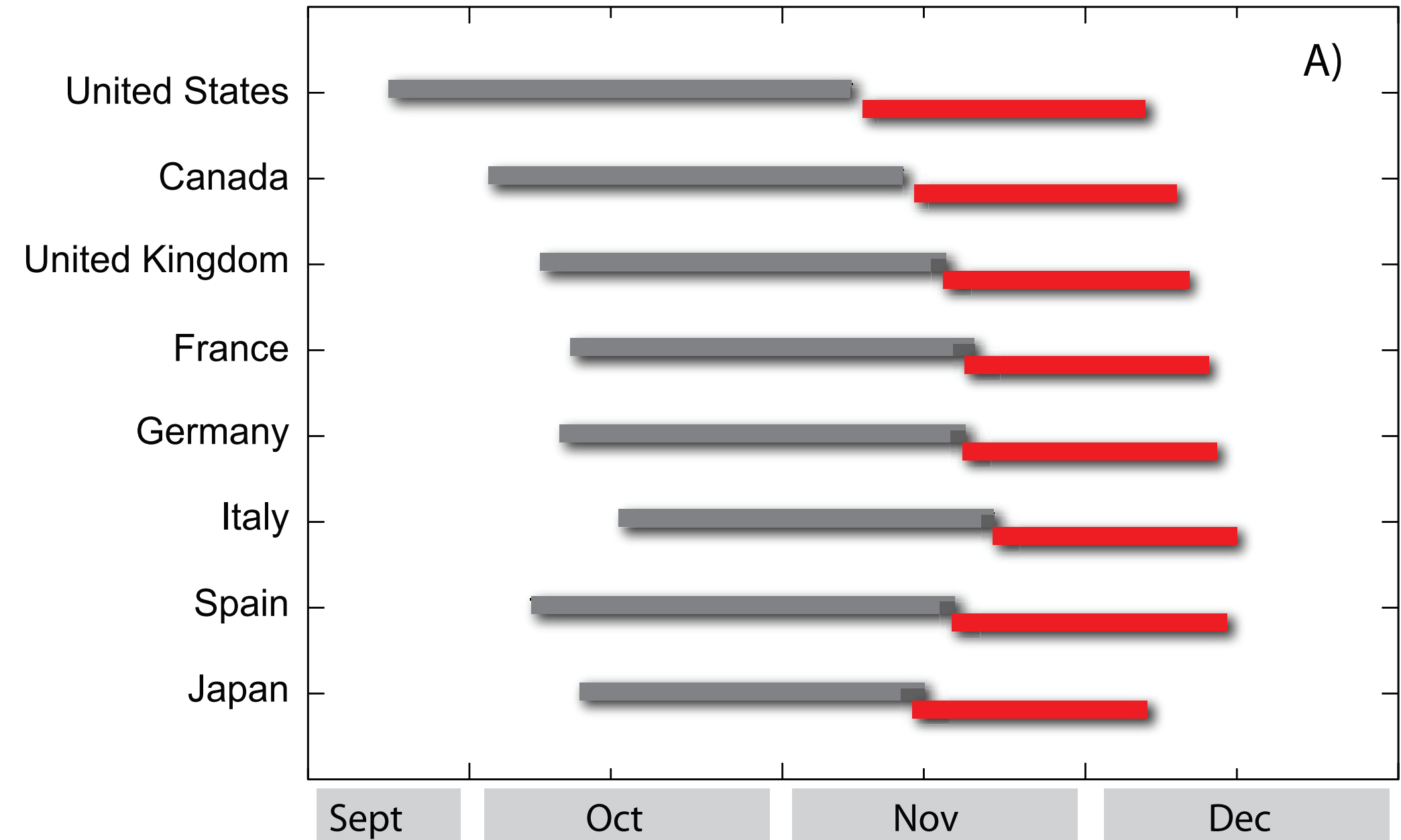

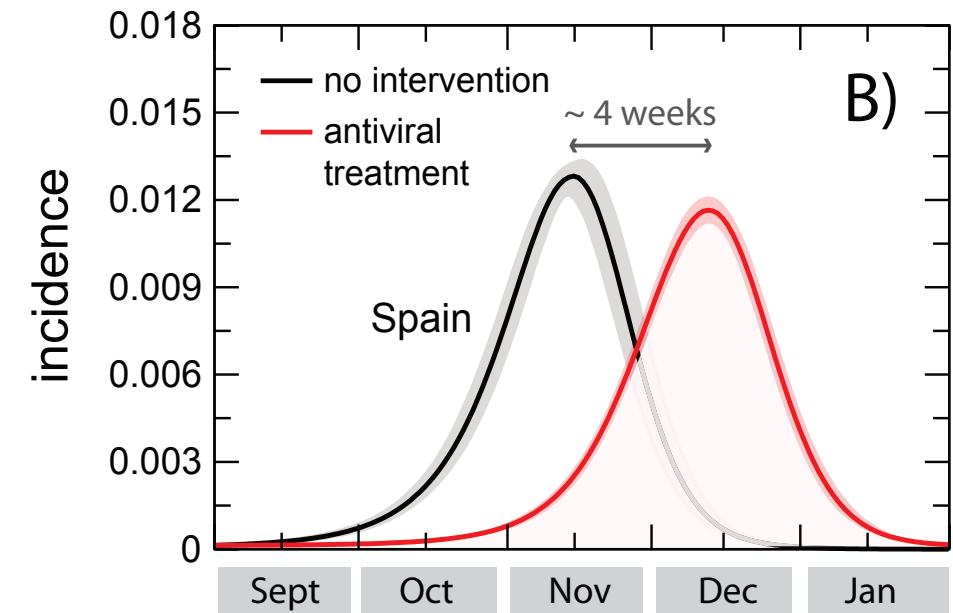

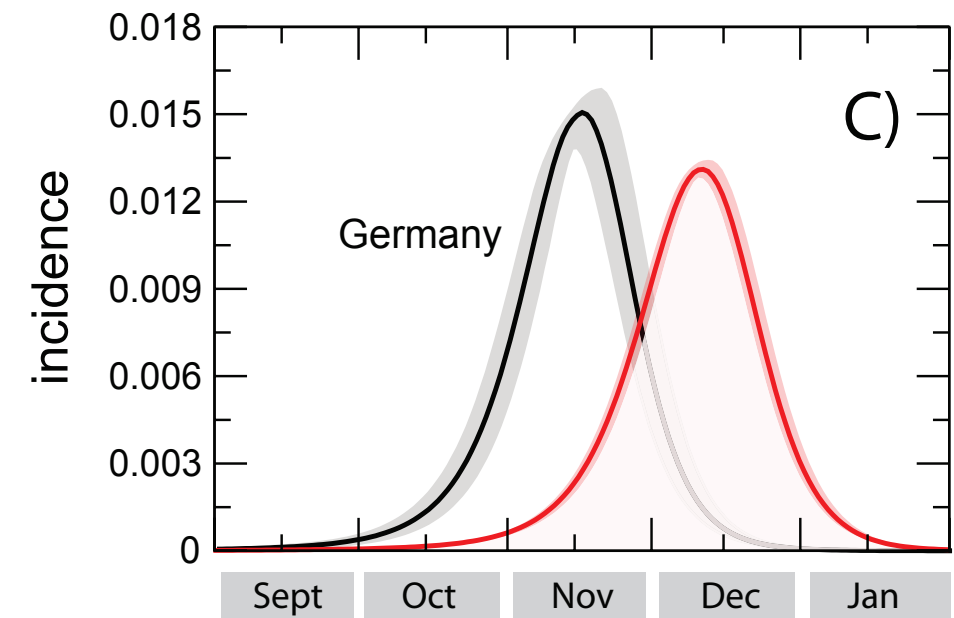

# Supplementary Information

*Seasonal transmission potential and activity peaks of the new influenza A(H1N1): a Monte Carlo likelihood analysis based on human mobility*

D Balcan, H Hu, B Gonçalves, P Bajardi, C Poletto,
JJ Ramasco, D Paolotti, N Perra, M Tizzoni, W Van den Broeck, V Colizza, A Vespignani

August 31, 2009

## 1 Structured metapopulation model

Here we present the detailed definition and data description of the global structured metapopulation model. The computational model is based on three data/model layers. The first layer is a data layer defining the census area and the subpopulation structure. The second one refers to human mobility model defined by the transportation and commuting networks characterizing the interactions and exchanges of individuals across subpopulations. The third layer is the epidemic dynamic model that defines the evolution of the infectious disease inside each subpopulations.

### 1.1 Global Population and subpopulations definition

The population dataset was obtained from the Web sites of the "Gridded Population of the World" and the "Global Urban-Rural Mapping" projects [1, 2], which are run by the Socioeconomic Data and Application Center (SEDAC) of Columbia University. The surface of the world is divided into a grid of cells that can have different resolution levels. Each of these cells has assigned an estimated population value.

Out of the possible resolutions, we have opted for cells of $15 \times 15$ minutes of arc to constitute the basis of our model. This corresponds to an area of each cell approximately equivalent to a rectangle of $25 \times 25$ kms along the Equator. The dataset comprises 823 680 cells, of which 250 206 are populated. Since the coordinates of each cell center and those of the WAN airports are known, the distance between the cells and the airports can be calculated. We have performed a Voronoi-like tessellation of the Earth surface assigning each cell to the closest airport that satisfies the following two conditions: (i) Each cell is assigned to the closest airport within the same country. And (ii), the distance between the airport and the cell cannot be longer than 200 kms. This cutoff naturally emerges from the distribution of distances between cells and closest airports, and it is introduced to avoid that in barely populated areas such as Siberia we can generate geographical census areas thousands of kilometer wide but with almost no population. It also corresponds to a reasonable upper cutoff for the ground traveling distance expected to be covered to reach an airport before traveling by plane.

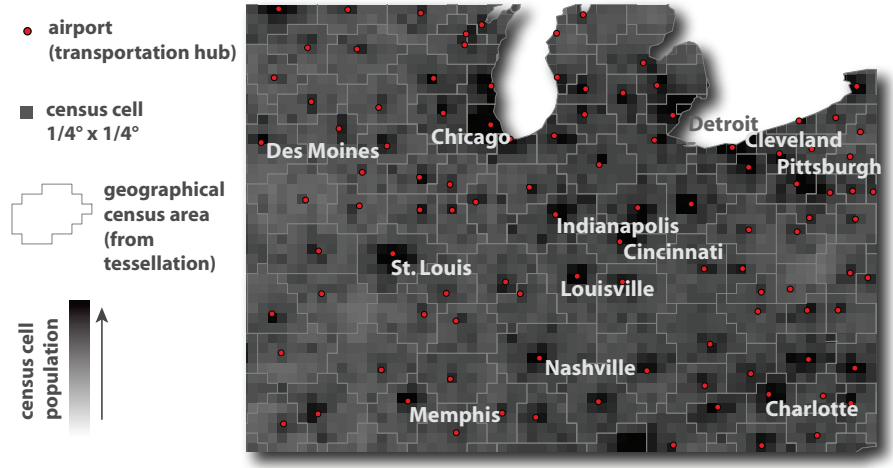

Figure 1: Population database and Voronoi tessellation around main tranportation hubs. The world surface is represented in a grid-like partition where each cell - corresponding to a population estimate - is assigned to the closest airport. Geographical census areas emerge that constitute the sub-populations of the meta-population model.

Before proceeding with the tessellation, we need to take into account that some urban areas include more than one airport. For instance, London has up to six airport, Paris has two, and New York City has three. Our aim is to build a metapopulation model whose subpopulations correspond to the geographical census areas obtained from tessellation. Inside these geographical census areas a homogeneous mixing is assumed. The groups of airports that serve the same urban area need therefore to be aggregated since the mixing within the given urban area is expected to be high and cannot be represented in terms of separated subpopulations for each of the airports serving the same city. We have searched for groups of airports located close to each other and we manually processed the identified groups of airports to select those belonging to the same urban area. The airports of the same group are then aggregated in a single "super-hub". An example with the final result of the Voronoi tessellation procedure with cells and airports can be seen in Figure 1. The geographical census areas become thus the basic subpopulations of our metapopulation model. Their connections will determine the geographical spreading of an hypothetical epidemic. The air transportation is already integrated in the model, but a further step must be taken in order to also include ground transportation in a realistic way.

Table 1: Commuting networks in each continent. Number of countries ($N_c$), number of administrative units (V) and inter-links between them (E) are summarized.

| Continent | $N_c$ | V | E |
|---|---|---|---|
| Europe | 17 | 65880 | 4490650 |
| North America | 2 | 6986 | 182255 |
| Latin America | 4 | 1858 | 63678 |
| Asia | 3 | 2732 | 323815 |
| Oceania | 2 | 746 | 30679 |
| Total | 28 | 78202 | 5091077 |

## 1.2  World Airport Network

The World Airport Network (WAN) is composed of 3362 commercial airports indexed by the International Air Transport Association (IATA) that are located in 220 different countries. The database contains the number of available seats per year for each direct connection between two of these airports. The coverage of the dataset is estimated to be 99% of the global commercial traffic. The WAN can be seen as a weighted graph comprising 16 846 edges whose weight, $\omega_{j\ell}$, represents the passenger flow between airports j and $\ell$. The network shows a high degree of heterogeneity both in the number of destinations per airport and in the number of passengers per connection [3, 4, 5, 6].

## 1.3  Commuting Networks

Our commuting databases have been collected from the Offices of Statistics of 28 countries in the 5 populated continents. The full dataset comprehends more than 78 000 administrative regions and over five million commuting flow connections between them (see [7]). The definition of administrative unit and the granularity level at which the commuting data are provided enormously vary from country to country. For example, most European countries adhere to a practice that ranks administrative divisions in terms of geocoding for statistical purposes, the so called Nomenclature of Territorial Units for Statistics (NUTS). Most countries in the European Union are partitioned into three NUTS levels which usually range from states to provinces. The commuting data at this level of resolution is therefore strongly coarse-grained. In order to have a higher geographical resolution of the commuting datasets that could match the resolution scale of our geographical census areas, we looked for smaller local administrative units (LAU) in Europe. The US or Canada report commuting at the level of counties. However, even within a single country the actual extension, shape, and population of the administrative divisions are usually a consequence of historical reasons and can be strongly heterogeneous.

Such heterogeneity renders the efforts to define a universal law describing commuting flows likely to fail. The mobility behavior might indeed result different across countries simply due to the country specific partition of the population into administrative boundaries. In order to over-come this problem, and in particular to define a data-driven short range commuting for GLEaM, we used the geographical census areas obtained from the Voronoi tessellation as the elementary units to define the centers of gravity for the process of commuting. This allows to deal with self-similar units across the world wth respect to mobility as emerged from a tessellation around main hubs of mobility and not country specific administrative boundaries.

We have therefore mapped the different levels of commuting data into the geographical census areas formed by the Voronoi-like tessellation procedure described above. The mapped commuting flows can be seen as a second transport network connecting subpopulations that are geographically close. This second network can be overlayed to the WAN in a multi-scale fashion to simulate realistic scenarios for disease spreading. The network exhibits important variability in the number of commuters on each connection as well as in the total number of commuters per geographical census area. Being the census areas relatively homogeneous and self-similar allows us to estimate a gravity law that successfully reproduce the commuting data obtained across different continents, and provide us with estimations for the possible commuting levels in the countries for which such data is not available as in ref. [7].

## 1.4   Epidemic dynamic model

Each geographical census area corresponds to a subpopulation in the metapopulation model, inside which we consider a Susceptible-Latent-Infectious-Recovered (SLIR) compartmental scheme, typical of influenza-like illnesses (ILIs), where each individual has a discrete disease state assigned at each moment in time. In Fig. 2, a diagram of the compartmental structure with transitions between compartments is shown. The contagion process, i.e. generation of new infections, is the only transition mechanism which is altered by short-range mobility, whereas all the other transitions between compartments are spontaneous and remain unaffected by the commuting. The rate at which a susceptible individual in subpopulation j acquires the infection, the so called force of infection $\lambda_j$, is determined by interactions with infectious persons either in the home subpopulation j or in its neighboring subpopulations on the commuting network.

Given the force of infection $\lambda_j$ in subpopulation j, each person in the susceptible compartment ($S_j$) contracts the infection with probability $\lambda_j \Delta t$ and enters the latent compartment ($L_j$), where $\Delta t$ is the time interval considered. Latent individuals exit the latent compartment with probability $\varepsilon \Delta t$, and transit to asymptomatic infectious compartment ($I_j^a$) with probability $p_a$ or, with the complementary probability $1 - p_a$, become symptomatic infectious. Infectious persons with symptoms are further divided between those who can travel ($I_j^t$), probability $p_t$, and those who are travel-restricted ($I_j^{nt}$) with probability $1 - p_t$. All the infectious persons permanently recover with probability $\mu \Delta t$, entering the recovered compartment ($R_j$) in the next time step. All transitions and corresponding rates are summarized in Table 2 and in Figure 2. In each subpopulation the variation of the number of individuals in each compartment [m] can be written at any given time step as

$$X_j^{[m]}(t + \Delta t) - X_j^{[m]}(t) = \Delta X_j^{[m]} + \Omega_j([m]) \qquad (1)$$

where the term $\Delta X_j^{[m]}$ represents the change due to the compartment transitions induced by the disease dynamics and the transport operator $\Omega_j([m])$ represents the variations due to the traveling and mobility of individuals. The latter operator takes into account the long-range airline mobility and define the minimal time scale of integration to 1 day. The mobility due to the commuting flows is taken into account by defining effective force of infections by using a time scale separation approximations as detailed in the following sections.

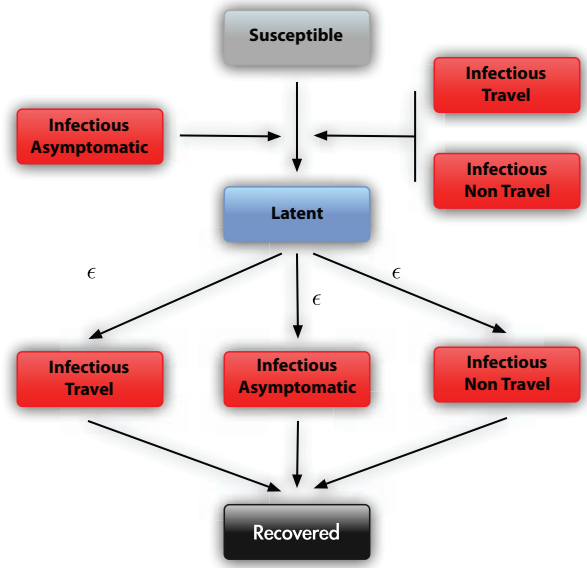

Figure 2: Compartmental structure of our epidemic model within each subpopulation.

The threshold parameter of the disease that determines the spreading rate of infection is called basic reproduction number ($R_0$), and is defined as the average number of infected cases generated by a typical infectious individual when introduced into a fully susceptible population [8]. For our compartmental model we have $R_0 = \beta \mu^{-1} [1 - p_a + r_\beta p_a]$.

## 1.5 Stochastic and discrete integration of the disease dynamics

In each subpopulation $j$, we define an operator acting on a compartment $[m]$ to account for all the transitions out of the compartment in the time interval $\Delta t$. Each element $\mathcal{D}_j([m], [n])$ of this operator is a random variable extracted from a multinomial distribution and determines the number of transitions from compartment $[m]$ to $[n]$ occurring in $\Delta t$. The change $\Delta X_j^{[m]}$ of a compartment $[m]$ in this time interval is given by a sum over all random variables $\{\mathcal{D}_j([m], [n])\}$ as follows

$$\Delta X_j^{[m]} = \sum_{[n]} \{ -\mathcal{D}_j([m], [n]) + \mathcal{D}_j([n], [m]) \} \quad . \tag{2}$$

As a concrete example let us consider the evolution of the latent compartment. There are three possible transitions from the compartment: transitions to the asymptomatic infectious, the symptomatic traveling and the non-traveling infectious compartments. The elements of the operator

Table 2: Transitions between compartments and their rates.

| Transition | Type | Rate |
|---|---|---|
| $S_j \to L_j$ | Contagion | $\lambda_j$ |
| $L_j \to I_j^a$ | Spontaneous | $\varepsilon p_a p_a$ |
| $L_j \to I_j^t$ | " | $\varepsilon(1 - p_a)p_t$ |
| $L_j \to I_j^{nt}$ | " | $\varepsilon(1 - p_a)(1 - p_t)$ |
| $I_j^a \to R_j$ | " | $\mu$ |
| $I_j^t \to R_j$ | " | $\mu$ |
| $I_j^{nt} \to R_j$ | " | $\mu$ |

acting on $L_j$ are extracted from the multinomial distribution

$$\mathrm{Pr}^{\mathrm{Multin}}(L_j(t), p_{L_j \to I_j^a}, p_{L_j \to I_j^t}, p_{L \to I_j^{nt}}) \qquad (3)$$

determined by the transition probabilities

$$\begin{aligned}
p_{L_j \to I_j^a} &= \varepsilon p_a \Delta t \quad, \\
p_{L_j \to I_j^t} &= \varepsilon(1 - p_a)p_t \Delta t \quad, \\
p_{L \to I_j^{nt}} &= \varepsilon(1 - p_a)(1 - p_t)\Delta t \quad,
\end{aligned} \qquad (4)$$

and by the number of individuals in the compartment $L_j(t)$ (its size). All these transitions cause a reduction in the size of the compartment. The increase in the compartment population is due to the transitions from susceptibles into latents. This is also a random number extracted from a binomial distribution

$$\mathrm{Pr}^{\mathrm{Bin}}(S_j(t), p_{S_j \to L_j}) \qquad (5)$$

given by the chance of contagion

$$p_{S_j \to L_j} = \lambda_j \Delta t \quad, \qquad (6)$$

with a number of attempts given by the number of susceptibles $S_j(t)$. After extracting these numbers from the appropriate multinomial distributions, we can calculate the change $\Delta L_j(t)$ as

$$\Delta L_j(t) = L_j(t + 1) - L_j(t) = -\left[\mathcal{D}_j(L, I^a) + \mathcal{D}_j(L, I^t) + \mathcal{D}_j(L, I^{nt})\right] + \mathcal{D}_j(S, L) \ . \qquad (7)$$

## 1.6  The integration of the transport operator

The transport operator is defined by the airline transportation data and sets the integration time scale to 1 day. The number of individuals in the compartment [m] traveling from the subpopulation j to the subpopulation $\ell$ is an integer random variable, in that each of the $X_j$ potential travelers has a probability $p_{j\ell} = w_{j\ell}/N_j$ to go from j to $\ell$. In each subpopulation j the numbers of individuals $\xi_{j\ell}$ traveling on each connection $j \rightarrow \ell$ at time t define a set of stochastic variables which follows the multinomial distribution

$$P(\{\xi_{j\ell}\}) = \frac{X_j^{[m]}!}{(X_j^{[m]} - \sum_\ell \xi_{j\ell})! \prod_\ell \xi_{j\ell}!} (1 - \sum_\ell p_{j\ell})^{(X_j^{[m]} - \sum_\ell \xi_{j\ell})} \prod_\ell p_{j\ell}^{\xi_{j\ell}}, \qquad (8)$$

where $(1 - \sum_\ell p_{j\ell})$ is the probability of not traveling, and $(X_j^{[m]} - \sum_\ell \xi_{j\ell})$ identifies the number of non traveling individuals of the compartment [m]. We use standard numerical subroutines to generate random numbers of travelers following these distributions. The transport operator in each subpopulation j is therefore written as

$$\Omega_j([m]) = \sum_\ell (\xi_{\ell j}(X_\ell^{[m]}) - \xi_{j\ell}(X_j^{[m]})), \qquad (9)$$

where the mean and variance of the stochastic variables are $\langle \xi_{j\ell}(X_j^{[m]}) \rangle = p_{j\ell} X_j^{[m]}$ and $\text{Var}(\xi_{j\ell}(X_j^{[m]})) = p_{j\ell}(1 - p_{j\ell})X_j^{[m]}$. It is worth remarking that on average the airline network flows are balanced so that the subpopulation $N_j$ are constant in time, e.g. $\sum_{[m]} \Omega_j([m]) = 0$. Direct flights as well as connecting flights up to two-legs flights can be considered. Connecting flights can be generated from the original IATA data by applying a transport operator that redistribute passengers flying on a one-step and two-steps flights, based on the fraction of passengers transiting at a given airport. Values collected for this parameter and the full description of the transport operator including connecting flights can be found in Ref. [5].

## 1.7   Time-scale separation and the integration of the commuting flows

The Global Epidemic and Mobility (GLEaM) modeler combines the infection dynamics with long- and short-range human mobility. Each of these dynamical processes operates at a different time scale. For ILI there are two important intrinsic time scales, given by the latency period $\varepsilon^{-1}$ and the duration of infectiousness $\mu^{-1}$, both larger than 1 day. The long-range mobility given by the airline network has a time scale of the order of 1 day, while the commuting takes place in a time scale of approx. $\tau^{-1} \sim 1/3$ day. The explicit implementation of the commuting in the model thus requires a time interval shorter than the minimal time of airline transportation. To overcome this problem, we use a time-scale separation technique, in which the short-time dynamics is integrated into an effective force of infection in each subpopulation.

We start by considering the temporal evolution of subpopulations linked only by commuting flows and evaluate the relaxation time to an equilibrium configuration. Consider the subpopulation j coupled by commuting to other n subpopulations. The commuting rate between the subpopulation j and each of its neighbors i will be given by $\sigma_{ji}$. The return rate of commuting individuals is set to be $\tau$. Following the work of Sattenspiel and Dietz [9], we can divide the individuals original from the subpopulation j, $N_j$, between $N_{jj}(t)$ who are from j are located in j at time t and those, $N_{ji}(t)$, that are from j are located in a neighboring subpopulation i at time t. Note that by consistency

$$N_j = N_{jj}(t) + \sum_i N_{ji}(t). \qquad (10)$$

The rate equations for the subpopulation size evolution are then

$$\partial_t N_{jj} = -\sum_i \sigma_{ji} N_{jj}(t) + \tau \sum_i N_{ji}(t) \quad,$$
$$\partial_t N_{ji} = \sigma_{ji} N_{jj}(t) - \tau N_{ji}(t) \quad.$$

(11)

By using condition (10), we can derive the closed expression

$$\partial_t N_{jj} + (\tau + \sigma_j) N_{jj}(t) = N_j \tau \quad,$$

(12)

where $\sigma_j$ denotes the total commuting rate of population j, $\sigma_j = \sum_i \sigma_{ji}$. $N_{jj}(t)$ can be expressed as

$$N_{jj}(t) = e^{-(\tau+\sigma_j)t} \left( C_{jj} + N_j \tau \int_0^t e^{(\tau+\sigma_j)s} ds \right) \quad,$$

(13)

where the constant $C_{jj}$ is determined from the initial conditions, $N_{jj}(0)$. The solution for $N_{jj}(t)$ is then

$$N_{jj}(t) = \frac{N_j}{(1 + \sigma_j/\tau)} + \left( N_{jj}(0) - \frac{N_j}{(1 + \sigma_j/\tau)} \right) e^{-\tau(1 + \sigma_j/\tau)t} \quad.$$

(14)

We can similarly solve the differential equation for the time evolution of $N_{ji}(t)$

$$\begin{aligned} N_{ji}(t) &= \frac{N_j \sigma_{ji}/\tau}{(1 + \sigma_j/\tau)} - \frac{\sigma_{ij}}{\sigma_j} \left( N_{jj}(0) - \frac{N_j}{(1 + \sigma_j/\tau)} \right) e^{-\tau(1 + \sigma_j/\tau)t} \\ &+ \left[ N_{ji}(0) - \frac{N_j \sigma_{ji}/\tau}{(1 + \sigma_j/\tau)} + \frac{\sigma_{ij}}{\sigma_j} \left( N_{jj}(0) - \frac{N_j}{(1 + \sigma_j/\tau)} \right) \right] e^{-\tau t} \quad. \end{aligned}$$

(15)

The relaxation to equilibrium of $N_{jj}$ and $N_{ji}$ is thus controlled by the characteristic time $[\tau(1 + \sigma_j/\tau)]^{-1}$ in the exponentials. Such term is dominated by $1/\tau$ if the relation $\tau \gg \sigma_j$ holds. In our case, $\sigma_j = \sum_i \omega_{ji}/N_j$, that equals the daily total rate of commuting for the population j. Such rate is always smaller than one since only a fraction of the local population is commuting, and it is typically much smaller than $\tau \simeq 3 - 10 \, \mathrm{day}^{-1}$. Therefore the relaxation characteristic time can be safely approximated by $1/\tau$. This time is considerably smaller than the typical time for the air connections of one day and hence our approximation of considering the subpopulations $N_{jj}(t)$ and $N_{ji}(t)$ as relaxed to their equilibrium values,

$$N_{jj} = \frac{N_j}{1 + \sigma_j/\tau} \quad \text{and} \quad N_{ji} = \frac{N_j \sigma_{ji}/\tau}{1 + \sigma_j/\tau} \quad,$$

(16)

is reasonable. This approximation, originally introduced by Keeling and Rohani [10], allows us to consider each subpopulation j as having an effective number of individuals $N_{ji}$ in contact with the individuals of the neighboring subpopulation i. In practice, this is similar to separate the commuting time scale from the other time scales in the problem (disease dynamics, traveling dynamics, etc.). While the approximation holds exactly only in the limit $\tau \to \infty$, it is good enough as long as $\tau^{-1}$ is much smaller than the typical transition rates of the disease dynamics. In the case of ILIs, the typical time scale separation between $\tau$ and the compartments transition rates is close to one order of magnitude or even larger. The Eqs. (16) can be then generalized in the time scale separation regime to all compartments [m] obtaining the general expression

$$X_{jj}^{[m]} = \frac{X_j^{[m]}}{(1 + \sigma_j/\tau)} \text{ and } X_{ji}^{[m]} = \frac{X_j^{[m]}}{(1 + \sigma_j/\tau)} \sigma_{ji}/\tau \,,$$  (17)

where $\sigma_j = \sum_{i \in \upsilon(j)} \sigma_{ji}$ denotes the total commuting rate of j. Whereas $X_{jj}^{[m]} = X_j^{[m]}$ and $X_{ji}^{[m]} = 0$ for all the other compartments which are restricted from traveling. These expressions will be used to obtain the effective force of infection taking into account the interactions generated by the commuting flows.

## 1.8   Effective force of infection

The force of infection $\lambda_j$ that a susceptible population of a subpopulation j sees can be decomposed into two terms: $\lambda_{jj}$ and $\lambda_{ji}$. The component $\lambda_{jj}$ refers to the part of the force of infection whose origin is local in j. While $\lambda_{ji}$ indicates the force of infection acting on susceptibles of j during their commuting travels to a neighboring subpopulation i. The effective force of infection can be estimated by summing these two terms weighted by the probabilities of finding a susceptible from j in the different locations, $S_{jj}/S_j$ and $S_{ji}/S_j$, respectively. Using the time-scale separation approximation that establishes the equilibrium populations in Eq. (16), we can write

$$\lambda_j = \frac{\lambda_{jj}}{1 + \sigma_j/\tau} + \sum_i \frac{\lambda_{ji}\sigma_{ji}/\tau}{1 + \sigma_j/\tau} \quad .$$  (18)

We will focus now on the calculation of each term of the previous expression. The force of infection occurring in a subpopulation j is due to the local infectious persons staying at j or to infectious individuals from a neighboring subpopulation i visiting j and so we can write

$$\lambda_{jj} = \frac{\beta_j}{N_j^*} \left[ I_{jj}^{nt} + I_{jj}^t + r_\beta I_{jj}^a + \sum_i \left( I_{ij}^{nt} + I_{ij}^t + r_\beta I_{ij}^a \right) \right] \quad ,$$  (19)

where $\beta_j$ is introduced to account for the seasonality in the infection transmission rate (if the seasonality is not considered, it is a constant), and $N_j^*$ stands for the total effective population in the subpopulation j. By definition, $I_{jj}^{nt} = I_j^{nt}$ and $I_{ji}^{nt} = 0$ for $j \neq i$. If we use the equilibrium values of the other infectious compartments (see Eq. (16)) we obtain

$$\lambda_{jj} = \frac{\beta_j}{N_j^*} \left[ I_j^{nt} + \frac{I_j^t + r_\beta I_j^a}{1 + \sigma_j/\tau} + \sum_i \frac{I_i^t + r_\beta I_i^a}{1 + \sigma_i/\tau} \sigma_{ij}/\tau \right] \quad .$$  (20)

The derivation of $\lambda_{ji}$ follows from a similar argument yielding:

$$\lambda_{ji} = \frac{\beta_i}{N_i^*} \left[ I_{ii}^{nt} + I_{ii}^t + r_\beta I_{ii}^a + \sum_{\ell \in \upsilon(i)} \left( I_{\ell i}^{nt} + I_{\ell i}^t + r_\beta I_{\ell i}^a \right) \right] \quad ,$$  (21)

where $\upsilon(i)$ represents the set of neighbors of $i$, and therefore the terms under the sum are due to the visits of infectious individuals from the subpopulations $\ell$, neighbors of $i$, to $i$. By plugging the equilibrium values of the compartment into the above expression, we obtain

$$\lambda_{ji} = \frac{\beta_i}{N_i^*}\left[I_i^{nt} + \frac{I_i^t + r_\beta I_i^a}{1 + \sigma_i/\tau} + \sum_{\ell \in \upsilon(i)} \frac{I_\ell^t + r_\beta I_\ell^a}{1 + \sigma_\ell/\tau}\sigma_{\ell i}/\tau\right] \quad . \tag{22}$$

Finally, in order to have an explicit form of the force of infection we need to evaluate the effective population size $N_j^*$ in each subpopulation $j$, i.e., the actual number of people actually staying at the location $j$. The effective population is $N_j^* = N_{jj} + \sum_i N_{ij}$, that in the time-scale separation approximation reads

$$N_j^* = I_j^{nt} + \frac{N_j - I_j^{nt}}{1 + \sigma_j/\tau} + \sum_i \frac{N_i - I_i^{nt}}{1 + \sigma_i/\tau}\sigma_{ij}/\tau \quad . \tag{23}$$

Note that in these equations all the terms with compartments have an implicit time dependence. By inserting $\lambda_{jj}$ and $\lambda_{ji}$ into Eq. (18), it can be seen that the expression for the force of infection includes terms of zeroth, first and second order on the commuting ratios (i.e., $\sigma_{ij}/\tau$). These three term types have a straightforward interpretation: The zeroth order terms represent the usual force of infection of the compartmental model with a single subpopulation. The first order terms account for the effective contribution generated by neighboring subpopulations with two different sources: Either susceptible individuals of subpopulation $j$ having contacts with infectious individuals of neighboring subpopulations $i$, or infectious individuals of subpopulations $i$ visiting subpopulation $j$. The second order terms correspond to an effective force of infection generated by the contacts of susceptible individuals of subpopulation $j$ meeting infectious individuals of subpopulation $\ell$ (neighbors of $i$) when both are visiting subpopulation $i$. This last term is very small in comparison with the zeroth and first order terms, typically around two order of magnitudes smaller, and in general can be neglected.

## 1.9   Seasonality modeling

In order to model the seasonality effect in the northern and in the southern hemispheres, we follow the approach of Cooper *et al* [15] rescaling the basic reproduction ratio $R_0$ by a sinusoidal function, $s_i(t)$,

$$s_i(t) = \frac{1}{2}\left[\left(1 - \frac{\alpha_{min}}{\alpha_{max}}\right)\sin\left(\frac{2\pi}{365}(t - t_{max,i}) + \frac{\pi}{2}\right) + 1 + \frac{\alpha_{min}}{\alpha_{max}}\right] \tag{24}$$

where $i$ refers to the hemisphere considered. In the tropical region the scaling function is identically equal to 1.0. $t_{max,i}$ is the time corresponding to the maximum of the sinusoid and hence to the maximum of the effective $R_0$, $R_{max} \equiv \alpha_{max}R_0$: it is fixed to Jan 15th in the northern hemisphere and six months later in the southern one. Along the year the seasonality scaling function varies from a maximum rescaling, $\alpha_{max}$, to a minimum rescaling $\alpha_{min}$. We have kept fixed $\alpha_{max}$ to 1.1 [16], corresponding to a slight increase of $R_0$, and tested different values of $\alpha_{min}$ from 0.1 to 0.9.

## 1.10   Control sanitary measures in Mexico

During the early stages of the epidemic, Mexican authorities implemented, under the supervision of WHO, a series of measures to increase social distance aimed at delaying the propagation of influenza. These measures consisted in school and college closure, suspension of acts involving people gathering such as concerts, masses, cinemas, etc, and the suspension for a few days of non essential economic activities. The effect of these measures has been a reduction in the number of cases reported between April 27th and May 10th and a consequent damping in the incidence curve that lasted at least until the beginning of June [17, 18].

The early stage evolution of the epidemic in Mexico is determinant for the world wide spread of the infection, therefore the authorities intervention has been taken in consideration as an ingredient of the model. We have simulated the social distances setting a low value of the basic reproductive ratio in Mexico, $R_0^{Mex}$, in the period between April 24th and May 10th, leaving out in this case the seasonality rescaling. In the baseline scenario we have set $R_0^{Mex} = 0.9$.

# 2   Likelihood analysis

We performed a Monte Carlo maximum-likelihood estimation [19, 20] of the epidemic basic reproductive number, $R_0$. We fit with our model the arrival dates, $t_i^*$, of the first infected individual in each country $i$. Due to the media attention, the country arrival times are the highest quality data among all the numbers regarding real epidemic cases and the least likely to be affected by under-reporting. Given our statistical model, the probability of the empirical set of arrival times $\{t_i^*\}$ conditioned to $R_0$, when seen as a function of $R_0$ itself, defines the likelihood function

$$\mathcal{L}(R_0) = \mathcal{P}(\{t_i^*\}|R_0).  \tag{25}$$

Maximizing this function, after fixing the values of the epidemiological and seasonality parameters ($\epsilon$, $\mu$, $\alpha_{min}$), we obtain an estimation of the basic reproductive number.

The probability distribution $\mathcal{P}(\{t_i\}|R_0)$ is embedded in the stochastic epidemic spreading process described by our model. Therefore we use a Monte Carlo methodology to sample the distribution by generating numerically $\{t_i\}$, for two thousand random realizations of the global epidemic spreading. Each stochastic run of the model starting with the same initial conditions and the same set of parameters ($\epsilon$, $\mu$, $\alpha_{min}$) yields a simulated arrival time $t_i$ of the first symptomatic case detected in a country not yet infected. Only symptomatic cases are considered, as asymptomatic individuals would go undetected. The set of 2000 numerical observations of the arrival times $t_i$ for each country $i$ allows the definition of the discrete probability $P_i(t_i)$ as the fraction of runs yielding arrival time $t_i$. This procedure iterated for different values of $R_0$ allows to reconstruct the likelihood function $\mathcal{L}(R_0)$.

In order to facilitate the sampling of the distribution we have restricted the set of arrival times, considering only the 12 countries (listed in Table 5) that have been seeded by Mexico both in the real epidemic and in the simulated process (at least in the 90% of the cases). Indeed the infection arrival dates in these countries are conditional statistical independent variables, being affected only by the epidemic evolution in Mexico. Therefore in this case we can factorize $\mathcal{P}(\{t_i\})$ in the

Table 3: Maximum-likelihood estimates of the basic reproductive number, $R_0$, varying the seasonality factor, $\alpha_{min}$, for the baseline scenario, $\epsilon^{-1} = 1.1$ and $\mu^{-1} = 2.5$.

| $\alpha_{min}$ | $R_0$ | [95%CI] |
|---|---|---|
| 0.1 | 1.76 | [1.65-1.87] |
| 0.3 | 1.76 | [1.65-1.87] |
| 0.5 | 1.78 | [1.66-1.90] |
| 0.6 | 1.76 | [1.64-1.88] |
| 0.7 | 1.76 | [1.64-1.88] |
| 0.8 | 1.76 | [1.65-1.87] |
| 0.9 | 1.75 | [1.63-1.87] |
| NS | 1.76 | [1.64-1.88] |

product of the arrival time distributions for each single country

$$\mathcal{P}\left(\{t_i\}\right) = \prod_i P_i(t_i). \qquad (26)$$

An example of $P_i(t_i)$ is shown in Fig. 3; it is a well peaked, smooth distribution, which allows us a good evaluation of the quantity $P_i(t_i^*)$.

In Table 3 we report the estimates of the $R_0$, and the corresponding confidence intervals, for the baseline scenario and different values of $\alpha_{min}$. Varying the seasonality scaling we obtain always the same value of $R_0$ within the errors. This result is due to the fact that almost the totality of Mexico is in the tropical region and thus seasonality does not affect the epidemic dynamic within this country and consequently neither the arrival times distribution. Therefore with this procedure we are able to leave out the seasonality parameter and obtain an independent estimation of the basic reproductive number.

## 3  Seasonality scaling

In order to obtain a rough estimation of the seasonality scaling factor, $\alpha_{min}$, we have analyzed the whole data set composed by the arrival dates of the first infected case in the 93 countries affected by the outbreak as of June 18th. This list of countries, reported in Table 6, extends the set of the 12 countries considered for the $R_0$ estimation in that we are interested in arrival times strongly affected by seasonality. Both in the real and in the simulated process, many of these countries are indeed seeded by the northern hemisphere where the dumping effect of seasonality had a determinant influence on the epidemic dynamic in the period from March to June spanned by our data.

We analyzed the correlation between the simulated arrival time by country and its corresponding empirical value, by measuring the regression coefficient (slope $\gamma$) between the two datasets.

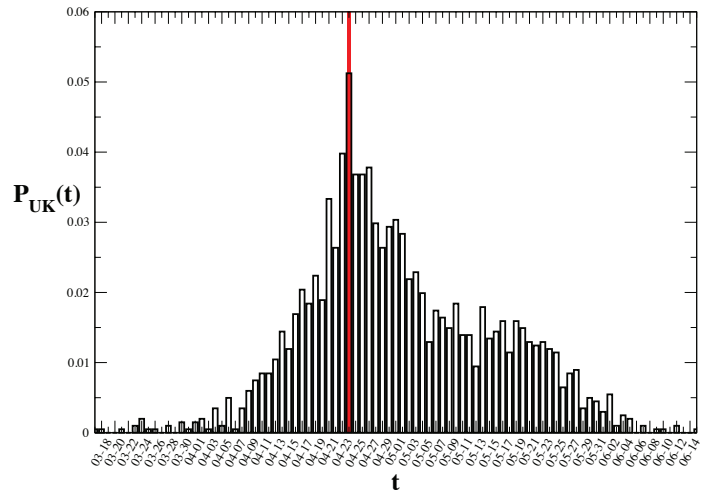

Figure 3: Distribution of the infection arrival times in the United States for the simulated spreading process corresponding to the baseline. The red bar marks the empirical data.

For the simulated arrival times we considered median and 95%CI. $\gamma$ was found to be sensitive to variations in the seasonality scaling factor, allowing the discrimination of the value of $\alpha_{min}$ that well fit the real epidemic, as the one having the slope closer to 1. For the baseline scenario, the best correlation was found with $\alpha_{min}$ between 0.6 and 0.7 (the corresponding plots are reported in Fig 4). Stronger and milder seasonalities lead respectively to a slower or faster epidemic pattern in respect to epidemic data.

# 4   Sensitivity analysis

Sensitivity analysis has been performed on the incubation and infectiousness periods, on the application of control measures in Mexico and on the initial date of the epidemic as well as on the empirical arrival date of infection in each country. In this section we illustrate all the different scenarios we have considered. In Table 4 we report the maximum likelihood values of $R_0$ with the 95%CI and the estimate of the $\alpha_{min}$ range obtained in each case, along with the resulting activity peak times in the different regions.

All the scenarios presented in the following sections contain variations with respect to the baseline case on one or more parameters, as follows. All the parameters are set to the baseline scenario if not stated otherwise.

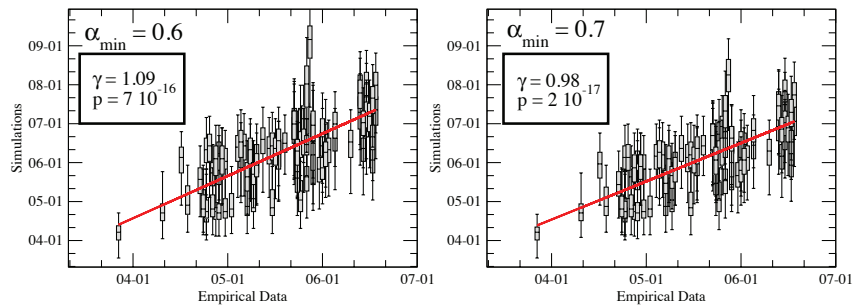

Figure 4: Simulated arrival times median and 95%CI versus the empirical ones for the two seasonality scaling factor 0.6 and 0.7.

Table 4: Results of the sensitivity analysis. Maximum likelihood value of $R_0$ and 95%CI, best seasonality scenarios, and predicted peak times in selected regions are shown.

| Best $R_0$ estimate and 95%CI | Range of $\alpha_{min}$ | World regions | Activity peak time [95%CI] |
|---|---|---|---|
| | | *Shorter infectious period, $\mu^{-1} = 1.1$ days* | |
| | | North America | [Jul 25-Oct 22] |
| | | Western Europe | [Aug 20-Nov 7] |
| 1.42 [1.34-1.50] | 0.8-0.9 | Southeast Asia | [Sept 11-Nov 9] |
| | | Upper South America | [Jul 28-Sept 15] |
| | | Lower South America | [Jul 28-Sept 13] |
| | | South Pacific | [Jul 22-Sept 20] |
| | | *Longer infectious period, $\mu^{-1} = 4$ days* | |
| | | North America | [Oct 19-Nov 15] |
| | | Western Europe | [Oct 31-Nov 24] |
| 2.11 [1.94-2.28] | 0.45-0.55 | Southeast Asia | [Sept 2-Oct 21] |
| | | Upper South America | [Jul 19-Aug 24] |

|  |  | Lower South America | [Jul 29-Aug 30] |
|  |  | South Pacific | [Jul 28-Sept 9] |
| *Longer latency period,* $\text{epsilon}^{-1} = 2.5$ *days* ($\mu^{-1} = 3$ *days*) | | | |
| 2.31 [2.14-2.48] | 0.4-0.5 | North America | [Oct 26-Nov 18] |
|  |  | Western Europe | [Nov 7-Nov 27] |
|  |  | Southeast Asia | [Sep 7-Oct 24] |
|  |  | Upper South America | [Jul 23-Aug 27] |
|  |  | Lower South America | [Aug 3-Sep 2] |
|  |  | South Pacific | [Aug 1-Sep 12] |
| *Anticipated arrival time* | | | |
| 1.87 [1.73-2.01] | 0.4-0.5 | North America | [Nov 13-Dec 7] |
|  |  | Western Europe | [Nov 17-Dec 6] |
|  |  | Southeast Asia | [Aug 16-Oct 4] |
|  |  | Upper South America | [Jul 4-Aug 4] |
|  |  | Lower South America | [Jul 17-Aug 15] |
|  |  | South Pacific | [Jul 15-Aug 24] |
| *Later initial date* | | | |
| 1.89 [1.77-2.01] | 0.45-0.55 | North America | [Oct 30-Nov 25] |
|  |  | Western Europe | [Nov 7-Nov 27] |
|  |  | Southeast Asia | [Aug 25-Oct 12] |
|  |  | Upper South America | [Jul 12-Aug 17] |
|  |  | Lower South America | [Jul 22-Aug 21] |
|  |  | South Pacific | [Jul 19-Sept 2] |
| *Earlier initial date* | | | |
| 1.65 [1.54-1.77] | 0.75-0.85 | North America | [Aug 3-Oct 18] |
|  |  | Western Europe | [Sept 1-Nov 6] |
|  |  | Southeast Asia | [Sept 16-Nov 13] |
|  |  | Upper South America | [Aug 1-Sept 17] |
|  |  | Lower South America | [Aug 4-Sept 18] |
|  |  | South Pacific | [Jul 28-Sept 25] |

## 4.1  Exploration of the model parameters

We explored different values of the disease parameters in order to gain insight into the variability in the optimal value of $R_0$ and $\alpha_{min}$ and the effects that different values can have on the simulations outcome. Three scenarios have been considered. In the first two cases we fixed the latency period to $\epsilon^{-1} = 1.1$ days and explored shorter and longer infectious period than the baseline value, with $\mu^{-1} = 1.1$ days and $\mu^{-1} = 4$ days, respectively. The latter case explored a longer latency periods than the baseline, $\epsilon^{-1} = 2.5$ days, by fixing the infectious period at $\mu^{-1} = 3$ days. All results show that longer generation times (as obtained by increasing the latency period or the infectious period) yield larger values of the maximum likelihood values of $R_0$.

A sensitivity on the value of $\alpha_{max}$ in the seasonality forcing (see the subsection 1.9) in the range [1.0-1.1] provides a maximum likelihood value for $R_0$ that does not differ from the baseline value within its confidence interval, yielding $R_0$=1.77 [95%CI: 1.66-1.88]. The range of seasonality scenarios is very close to the one obtained for the baseline, with $\alpha_{min}$ in [0.65-0.75]. The studied variation in the winter rescaling factor therefore does not affect the baseline scenario.

We also studied variations in the relative infectiousness of asymptomatic individuals, $r_\beta$, from 20% to 80%, finding a maximum likelihood value of $R_0$ that is within the confidence interval of the baseline value in both cases. The obtained values are $R_0$=1.77 [95%CI: 1.65-1.89] for $r_\beta$=20%, and $R_0$=1.77 [95%CI: 1.63-1.91] for $r_\beta$=80%.

## 4.2   Control sanitary measures in Mexico

As part of the sensitivity analysis, we have tested different values of $R_0^{Mex}$ (see section 1.10). Values of $R^{Mex}$ between 0.5 and 1.2 [18] have been tested, with no relevant changes. The scenario in which no sanitary measures have been applied, i.e. $R_0^{Mex} \equiv R_0$ was also explored. While the maximum likelihood value of $R_0$ does not change within the error bars, yielding $R_0$=1.75 [95%CI: 1.64-1.86], $\alpha_{min}$ is decreased and is found in the range [0.45-0.55]. This result can be explained considering that most of the 93 countries taken into account for the linear regression analysis are seeded either from US or from Mexico. The detection of the seasonal effect relies on an interplay between the number of cases exported by each of these two countries. With respect to the baseline scenario, a higher effective $R_0$ in Mexico determines more countries seeded from this country and less from the US, which occurs with an enhanced seasonality.

## 4.3   Uncertainties in the chronological data

To assess the effects of the uncertainties in the chronological data, we have tested three different scenarios. We have first anticipated of one week the empirical data of the infection arrival time in each country. We consider indeed one week as the worst possible delay in the first case detection. The shift of $-7$ days raised the best value of $R_0$ to 1.87, a value that is still within the confidence interval of the maximum likelihood value of the baseline case (see Table 3 for comparison).

In order to understand the effects of variations in the beginning of the epidemic, we have simulated the scenarios in which the spreading starts one week later and one week earlier with respect to the date provided by official sources and used in the baseline scenario. February 25th and February 11th have been set as initial dates, respectively, in the two cases. As expected, the delay of one week in the pandemic initial date results in a higher value of the reproductive number with respect to the baseline scenario, with $R_0$ =1.89 [95%CI: 1.77-2.01]. On the other hand, the anticipation of the start of the epidemic results in lowering the value of $R_0$ to 1.65 [95%CI: 1.54-1.77], and moving the best values of $\alpha_{min}$ to $0.75 - 0.85$.

# 5   Chronology and data on the H1N1 worldwide spreading

In the table (5) the onset of symptoms, flight arrival and official confirmation dates are presented. We focused on 12 countries seeded by Mexico for which we could find a clear description of the first confirmed case. When available, official data and reports were used. When the official sources of information on a certain country were incomplete, we relied on news from the local press.

Table 5: Dates for the onset of symptoms, flight arrival and official confirmation of the first confirmed case in 12 countries seeded by Mexico are reported.

| Country | Onset of symptoms | Flight arrival | Confirmed on |
|---|---|---|---|
| United States | March 28 [21] | – | April 21 [21] |
| Canada | April 11 [22] | April 8 [23] | April 23 [24] |
| El Salvador | – | April 19 [25] | May 3 [26] |
| United Kingdom | April 24 [27] | April 21 [28] | April 27 [24] |
| Spain | April 25 [29] | April 22 [30] | April 27 [24] |
| Cuba | – | April 25 [31] | May 13 [24] |
| Costa Rica | April 25 [32] | April 25 [32] | May 2 [24] |
| Netherlands | – | April 27 [33] | April 30 [33] |
| Germany | April 28 [34] | – | April 29 [24] |
| France | – | – | May 1 [35] |
| Guatemala | May 1 [36] | – | May 5 [37] |
| Colombia | – | – | May 3 [38] |

In the table (6) we show the complete timeline of confirmed cases for countries with at least one case as of July 18. The reported date corresponds to the onset of symptoms. For those countries where this information was not available, we report the flight arrival date or, when even this information was not available, we display the official confirmation date.

Table 6: Timeline of the confirmed cases as of June 18

| Country | Confirmation | Ref. | Country | Confirmation | Ref. |
|---|---|---|---|---|---|
| Israel | April 26 | [39] | New Zealand | April 25 | [40] |
| Austria | April 17 | [41] | Switzerland | April 29 | [42] |
| Ireland | April 30 | [43] | Denmark | April 29 | [44] |
| Hong Kong | April 30 | [45] | Italy | April 23 | [46] |
| South Korea | April 28 | [47] | Portugal | April 30 | [48] |
| Sweden | May 6 | [49] | Argentina | April 27 | [50] |
| Poland | May 5 | [51] | Japan | May 8 | [52] |
| Panama | April 24 | [53] | Brazil | May 8 | [49] |
| Norway | May 6 | [54] | Australia | May 18 | [55] |
| Finland | May 6 | [56] | Mainland China | May 9 | [57] |

| Thailand | May 9 | [58] | Belgium | May 11 | [59] |
|---|---|---|---|---|---|
| Ecuador | May 10 | [60] | Peru | May 10 | [61] |
| Malaysia | May 13 | [62] | Turkey | May 14 | [63] |
| Chile | May 16 | [64] | Honduras | May 11 | [65] |
| India | May 16 | [49] | Greece | May 17 | [66] |
| Philippines | May 18 | [67] | Russian Federation | May 18 | [68] |
| Paraguay | May 20 | [69] | Iceland | May 23 | [70] |
| Kuwait | May 23 | [71] | Lebanon | May 23 | [72] |
| Romania | May 23 | [73] | Bolivia | May 24 | [74] |
| United Arab Emirates | May 24 | [75] | Puerto Rico | May 24 | [76] |
| Singapore | May 24 | [77] | Bahrain | May 25 | [78] |
| Czech Republic | May 25 | [79] | Venezuela | May 25 | [80] |
| Hungary | May 26 | [81] | Vietnam | May 26 | [82] |
| Bulgaria | May 27 | [83] | Dominican Republic | May 27 | [84] |
| Estonia | May 27 | [85] | Uruguay | May 27 | [86] |
| Slovakia | May 28 | [87] | Bahamas | May 29 | [88] |
| Cayman Islands | May 29 | [89] | Cyprus | May 29 | [90] |
| Saudi Arabia | May 29 | [91] | Ukraine | May 29 | [92] |
| Jamaica | May 31 | [93] | Egypt | June 1 | [94] |
| Luxembourg | June 1 | [95] | Nicaragua | June 1 | [96] |
| Bermuda | June 2 | [97] | Trinidad and Tobago | May 30 | [98] |
| Barbados | June 3 | [99] | Dominica | June 5 | [100] |
| French Polynesia | June 5 | [101] | Morocco | June 10 | [102] |
| Yemen | June 13 | [103] | Oman | June 13 | [104] |
| Qatar | June 13 | [105] | South Africa | June 14 | [106] |
| Suriname | June 14 | [107] | Papua New Guinea | June 15 | [108] |
| Jordan | June 15 | [109] | Western Samoa | June 15 | [110] |
| Solomon Islands | June 15 | [111] | Sri Lanka | June 16 | [112] |
| British Virgin Islands | June 17 | [113] | Macau | June 17 | [114] |
| Netherlands Antilles | June 17 | [115] | Bangladesh | June 18 | [116] |
| Lao Peoples Rep | June 18 | [117] | | | |

# 6   Hospitalization

The ratio between hospitalized individual and confirmed cases is probably an overestimation of the hospitalization rate (HR) because of the under-ascertainment of the real infected individuals. Furthermore, the HR shows large fluctuations from one country to another due to different monitoring systems and containment policies. As a baseline, we consider an HR=10% [118], as calculated from confirmed cases for the United States. According to ref [119] a more realistic HR can be estimated using the multiplier method. Since the number of infected individuals has been estimated to be 10-30 times larger than laboratory-confirmed cases, the HR can be calculated from real data by taking into account this scaling factor. In table 3 of the main paper values obtained for these HR are shown and compared with the one for a typical seasonal flu [120] [121].

Table 7: Weekly new cases (thousands): comparison between reported cases and simulation results.

| Week | Simulated [95%CI] | Reported | [10x − 30x] |
|------|-------------------|----------|-------------|
| *USA* | | | |
| May 10 - May 16 | [0.03-11.91] | 2.46 | [26.60-73.80] |
| Jun 7 - Jun 13 | [1.32-152.48] | 4.64 | [46.38-139.14] |
| Jul 5 - Jul 11 | [32.16-1031.36] | 3.26 | [32.63-97.89] |
| *UK* | | | |
| May 10 - May 16 | [0.00-0.32] | 0.05 | [0.53-1.59] |
| Jun 7 - Jun 13 | [0.02-4.35] | 0.59 | [5.93-17.79] |
| Jul 5 - Jul 11 | [0.49-29.74] | 2.29 | [22.93-68.79] |
| *AUSTRALIA* | | | |
| May 10 - May 16 | [0.00-2.84] | 0.42 | [4.21-12.63] |
| Jun 7 - Jun 13 | [0.00-151.91] | 1.58 | [15.77-47.31] |
| Jul 5 - Jul 11 | [0.28-2263.80] | 4.58 | [45.84-137.52] |

# 7    Comparison of simulation results and real data

We compare the simulation results obtained for the baseline case with the reported data of a selection of countries. Table 7 shows the weekly new number of cases (expressed in thousands) for a set of three dates from early May to the end of July, for the United States, the United Kingdom, and Australia. The ranges in the simulated results correspond to the 95% CI obtained from the baseline simulations. Along with the number of reported cases, obtained from official sources [123, 124, 125, 126], we also report an interval calculated assuming a multiplier method of 10 and 30 times the confirmed cases. Results show an enhanced surveillance during the early phase of the outbreak, followed by a progressive decrease of the monitoring capacity. The larger number of cases made it indeed increasingly difficult to closely monitor the epidemic by confirming cases, and reporting requirements changed for countries that experienced an early outbreak [122].

Figure 5 reports the comparison of the epidemic evolution in 4 Territories of Australia – Western Australia, New South Wales, Queensland, and Northern Territory. The cumulative number of reported cases by Territory [126] (red circles) is compared to the corresponding predicted values, as indicated by the 95% CI (gray area). The plots show a good agreement between the empirical curves and the predicted behaviors, with the reported data being within the confidence interval of the simulations. Most importantly, the simulations are able to reproduce the heterogeneous geo-temporal pattern observed in the Australian Territories, where the outbreak started at different times. As an example, we report the time at which the first cluster of cases was detected in New South Wales and in Queensland, as indicated by the vertical dashed lines. The delay of approximately 2 weeks between the two outbreaks is correctly predicted by the simulations that take into account the mobility of individuals within the country (from actual commuting data) and with abroad.

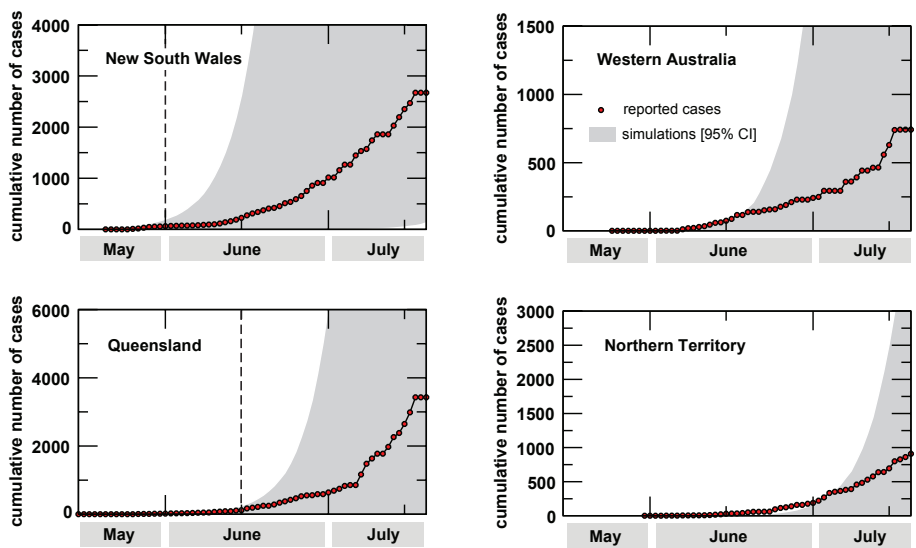

Figure 5: Comparison of reported (red circles) vs. simulated cases (95%CI indicated by the gray area) for a selection of 4 Territories of Australia. The simulations correctly predicts the different temporal evolutions observed in distinct Territories, as induced by the mobility of individuals. The vertical dashed lines indicate the delay of the start of the outbreak observed in two different Territories.